\documentclass{article}
\pdfoutput=1
\usepackage[T1]{fontenc}
\usepackage[latin9]{inputenc}
\usepackage{geometry}
\usepackage{verbatim}
\usepackage{amsmath}
\usepackage{amssymb}
\usepackage{graphicx}

\makeatletter

\let\jnl@style=\rm
\def\ref@jnl#1{{\jnl@style#1}}

\def\aj{\ref@jnl{AJ}}                   
\def\actaa{\ref@jnl{Acta Astron.}}      
\def\araa{\ref@jnl{ARA\&A}}             
\def\apj{\ref@jnl{ApJ}}                 
\def\apjl{\ref@jnl{ApJ}}                
\def\apjs{\ref@jnl{ApJS}}               
\def\ao{\ref@jnl{Appl.~Opt.}}           
\def\apss{\ref@jnl{Ap\&SS}}             
\def\aap{\ref@jnl{A\&A}}                
\def\aapr{\ref@jnl{A\&A~Rev.}}          
\def\aaps{\ref@jnl{A\&AS}}              
\def\azh{\ref@jnl{AZh}}                 
\def\baas{\ref@jnl{BAAS}}               
\def\bac{\ref@jnl{Bull. astr. Inst. Czechosl.}}
\def\caa{\ref@jnl{Chinese Astron. Astrophys.}}
\def\cjaa{\ref@jnl{Chinese J. Astron. Astrophys.}}
\def\icarus{\ref@jnl{Icarus}}           
\def\jcap{\ref@jnl{J. Cosmology Astropart. Phys.}}
\def\jrasc{\ref@jnl{JRASC}}             
\def\memras{\ref@jnl{MmRAS}}            
\def\mnras{\ref@jnl{MNRAS}}             
\def\na{\ref@jnl{New A}}                
\def\nar{\ref@jnl{New A Rev.}}          
\def\pra{\ref@jnl{Phys.~Rev.~A}}        
\def\prb{\ref@jnl{Phys.~Rev.~B}}        
\def\prc{\ref@jnl{Phys.~Rev.~C}}        
\def\prd{\ref@jnl{Phys.~Rev.~D}}        
\def\pre{\ref@jnl{Phys.~Rev.~E}}        
\def\prl{\ref@jnl{Phys.~Rev.~Lett.}}    
\def\pasa{\ref@jnl{PASA}}               
\def\pasp{\ref@jnl{PASP}}               
\def\pasj{\ref@jnl{PASJ}}               
\def\rmxaa{\ref@jnl{Rev. Mexicana Astron. Astrofis.}}%
\def\qjras{\ref@jnl{QJRAS}}             
\def\skytel{\ref@jnl{S\&T}}             
\def\solphys{\ref@jnl{Sol.~Phys.}}      
\def\sovast{\ref@jnl{Soviet~Ast.}}      
\def\ssr{\ref@jnl{Space~Sci.~Rev.}}     
\def\zap{\ref@jnl{ZAp}}                 
\def\nat{\ref@jnl{Nature}}              
\def\iaucirc{\ref@jnl{IAU~Circ.}}       
\def\aplett{\ref@jnl{Astrophys.~Lett.}} 
\def\apspr{\ref@jnl{Astrophys.~Space~Phys.~Res.}}
\def\bain{\ref@jnl{Bull.~Astron.~Inst.~Netherlands}} 
\def\fcp{\ref@jnl{Fund.~Cosmic~Phys.}}  
\def\gca{\ref@jnl{Geochim.~Cosmochim.~Acta}}   
\def\grl{\ref@jnl{Geophys.~Res.~Lett.}} 
\def\jcp{\ref@jnl{J.~Chem.~Phys.}}      
\def\jgr{\ref@jnl{J.~Geophys.~Res.}}    
\def\jqsrt{\ref@jnl{J.~Quant.~Spec.~Radiat.~Transf.}}
\def\memsai{\ref@jnl{Mem.~Soc.~Astron.~Italiana}}
\def\nphysa{\ref@jnl{Nucl.~Phys.~A}}   
\def\physrep{\ref@jnl{Phys.~Rep.}}   
\def\physscr{\ref@jnl{Phys.~Scr}}   
\def\planss{\ref@jnl{Planet.~Space~Sci.}}   
\def\procspie{\ref@jnl{Proc.~SPIE}}   

\makeatother

\begin{document}

\title{A preceding low-virulence strain pandemic \\
inducing immunity against COVID-19}

\author{Hagai B. Perets$^{1}$\thanks{Corresponding author: hperets@physics.technion.ac.il}$\,$
and Ruth Perets$^{1,2}$}
\maketitle
\begin{center}
$^{1}$Technion - Israel Institute of Technology, Haifa, Israel 3200004\\
$^{2}$Rambam Health Care Campus, Haifa, Israel 
\par\end{center}
\begin{abstract}
The COVID-19 pandemic is thought to began in Wuhan, China in December
2019. Mobility analysis identified East-Asia and Oceania countries
to be highly-exposed to COVID-19 spread, consistent with the earliest
spread occurring in these regions. However, here we show that while
a strong positive correlation between case-numbers and exposure level
could be seen early-on as expected, at later times the infection-level
is found to be \emph{negatively} correlated with exposure-level. Moreover,
the infection level is positively correlated with the population size,
which is puzzling since it has not reached the level necessary for
population-size to affect infection-level through herd immunity. These
issues are resolved if a low-virulence Corona-strain (LVS) began spreading
earlier in China outside of Wuhan, and later globally, providing immunity
from the later appearing high-virulence strain (HVS). Following its
spread into Wuhan, cumulative mutations gave rise to the emergence
of a HVS, known as SARS-CoV-2, starting the COVID-19 pandemic. We
model the co-infection by a LVS and a HVS, and show that it can explain
the evolution of the COVID-19 pandemic and the non-trivial dependence
on the exposure level to China and the population-size in each country.
We find that the LVS began its spread a few months before the onset
of the HVS, and that its spread doubling-time is $\sim1.59\pm0.17$
times slower than the HVS. Although more slowly spreading, its earlier
onset allowed the LVS to spread globally before the emergence of the
HVS. In particular, in countries exposed earlier to the LVS and/or
having smaller population-size, the LVS could achieve herd-immunity
earlier, and quench the later-spread HVS at earlier stages. We find
our two-parameter (the spread-rate and the initial onset time of the
LVS) can accurately explain the current infection levels (${\rm R^{2}=0.74}$;
correlation p-value (p) of ${\rm 5.2\times10^{-13}}$). Furthermore,
countries exposed early should have already achieved herd-immunity.
We predict that in those countries cumulative infection levels could
rise by no more than 2-3 times the current level through local-outbreaks,
even in the absence of any containment measures. We suggest several
tests and predictions to further verify the double-strain co-infection
model, and discuss the implications of identifying the LVS.
\end{abstract}
\begin{figure}
\includegraphics[scale=0.45]{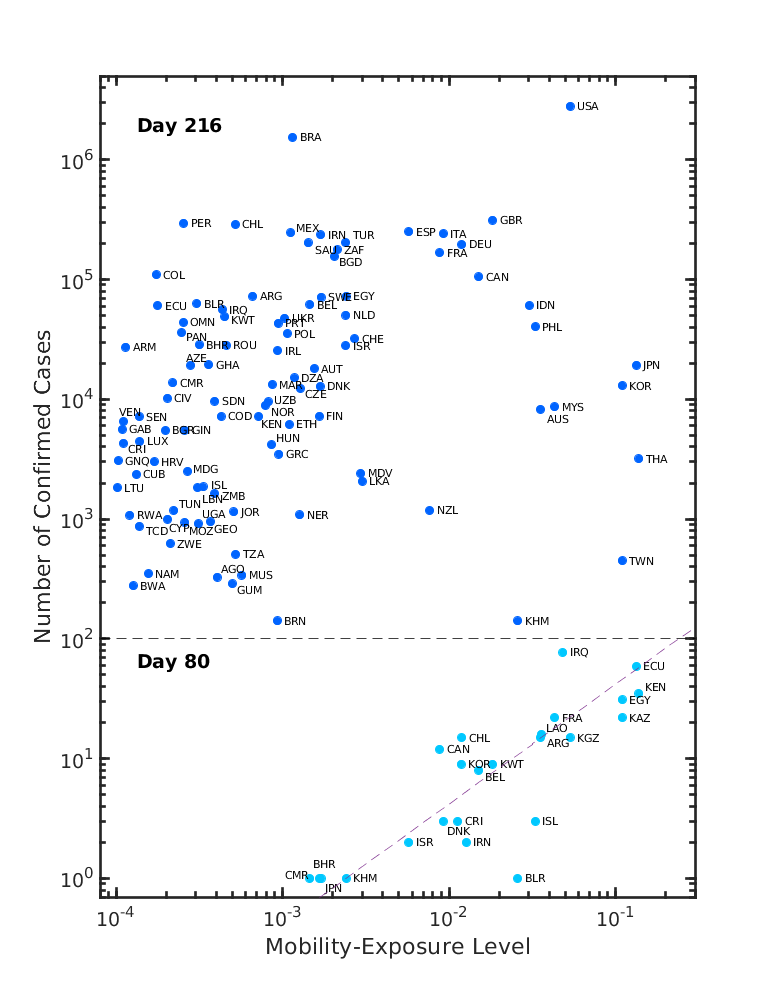}\includegraphics[scale=0.45]{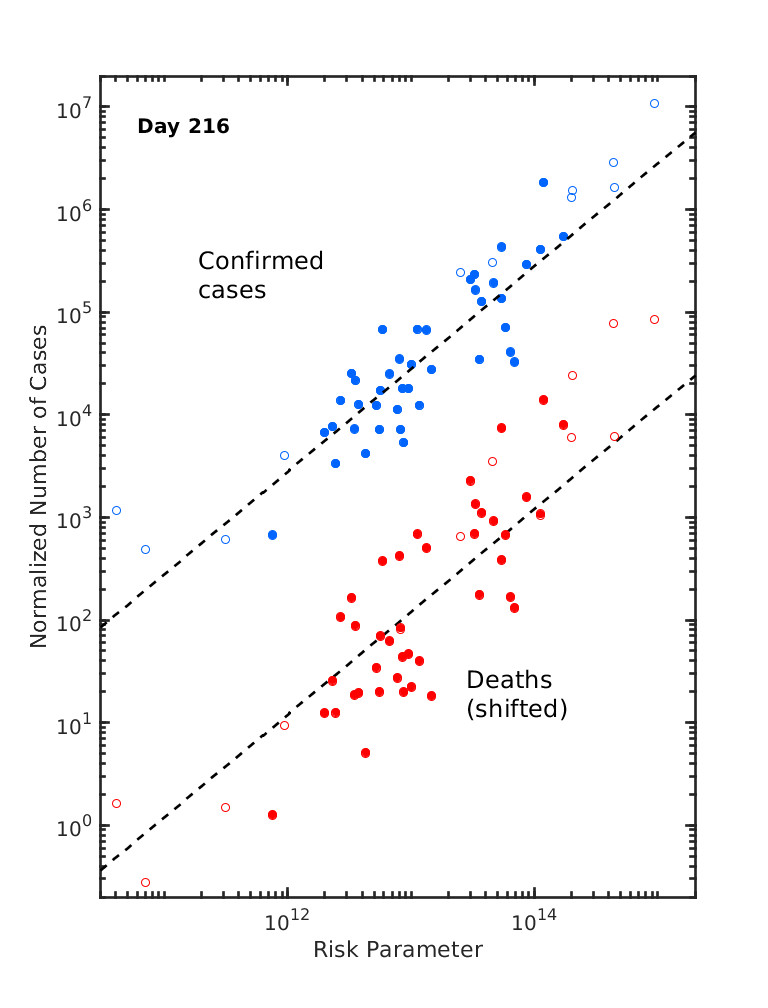}

\caption{\label{fig:Correlation}Left: Comparison of the number of confirmed
cases as a function of the exposure level, at early (80 days since
December 1$^{{\rm st}}$2019) and late (216 days) times. Early on
highly exposed countries are infected earlier, and show an increasing
number of cases as a function of the mobility-exposure level. At late
times the same countries show a \emph{negative} correlation between
the infection-level and the mobility-exposure level. Right: The testing-normalized
numbers of confirmed cases and testing (and age) normalized numbers
of deaths (shifted down by a factor of ten for clarity), as a function
of the risk-parameter (see definition in the main text) showing a
direct linear relation as expected in the double-strain co-infection
model. The least-populated and highest-populated countries (marked
by open circles) were not used in deriving the models parameters as
to assure they do not bias the results (see methods). Nevertheless,
as can be seen, these too are also consistent with the model. See
methods for a similar figure with annotated countries names (not shown
here for clarity).}
\end{figure}

COVID-19 is thought to be a zoonotic disease, which currently spreads
by human-to-human transmission\cite{Lu+20,Wu+20}. Human mobility
mediates the geographic spread of the disease between countries through
ground and air transportation. Given the origin of COVID-19 in China,
it is therefore expected that traffic from China to other countries
would drive the initial epidemic spread world-wide (e.g. \cite{Hai+20}).
In particular, it is expected that outbound mobility levels from China
into other countries (i.e. the numbers of incoming passengers from
China) should predict the risk/exposure level of the disease-spread
into these regions. Here we make use of exposure-level, defined as
the mobility level and risk assessment provided by the GLEAM epirisk
module (e.g. Refs. \cite{Van+11,Pas+19,Chi+20}), and compare it with
the infection level (see Fig. \ref{fig:Correlation}a), measured either
by the number of confirmed COVID-19 cases or confirmed COVID-19 deaths
(and later accounting for normalization by testing levels and/or age-structure;
see Methods).

We find that shortly after the beginning of COVID-19 spread, highly
exposed countries show a linear positive correlation between the infection
level (as measured by the number of COVID-19 confirmed cases), and
the mobility-exposure level (given by the epiRisk module, based on
flight-traffic data) at early times (${\rm p=5.5\times10^{-8}};$
for countries with first cases by 80 days after December 1st 2019;
see additional discussion of the early spread in Methods), but show
no correlation with the population size \footnote{Note that unless otherwise stated, all correlations discussed are
of the logarithms of the parameters}. However, at a later time (day 216), the same countries show a negative
correlation between the infection-level and the mobility-exposure
level (${\rm p}=0.014$, $\rho=-0.66$; and ${\rm p=0.006}$; $\rho=-0.7$
, for age-normalized deaths), as can be seen in Fig. \ref{fig:Correlation}a.
Moreover, we find that at the later time point, the infection-level
for all countries in our sample is positively correlated with the
population-size (Pearson test; ${\rm p}=1.2\times10^{-9}$ $\rho=0.55$
for confirmed cases; and ${\rm {\rm p}=2.8\times10^{-13}}$, $\rho=0.63$
for age-normalized deaths). While the early direct correlation of
cases with exposure supports exposure-level risk estimates, our attention
was drown to the unexpected negative correlation between the infection-level
and and exposure-level at later times, and the correlation with population
size at a relatively early epidemic stage where the numbers of cases
are far below the levels necessary for herd immunity to take affect. 

Here we show that the existence of an earlier low virulence (ancestor)
strain (LVS) providing immunity from the HVS, could explain the current
global distribution of the COVID-19. We propose that the LVS spread
from a location in China, but far outside Wuhan, and later gave rise
to the emergence of a the highly virulent strain (HVS), known as SARS-CoV-2.
The exact location of the initial LVS's outbreak is unknown, but is
likely to be in one of the least HVS-infected regions in China, and
possibly close to Vietnam, given the very low infection levels observed
there (see also section \ref{subsec: systematics}).

Previous studies explored epidemic models of co-infection and cross-immunity
by multiple viral strains\cite{Die+79,Cas+89,Adl+91,Bal+11,Sus+15,Kuc+16,Nic+16,Tho+19,Kam+20}.
The emergence of highly pathogenic respiratory viruses from low-pathogenic
ancestors has been previously described\cite{Qi+2018}, and cross-immunity
between respiratory low-virulence strains and high-virulence strains
has been described as well\cite{Seo+01,Kha+09,Nfo+12}. In these cases
the LVS effectively gave rise to a natural live vaccine, leading to
a mild-to-asymptomatic presentation of the HVS disease later-on \cite{Seo+01,Kha+09,Nfo+12}.
In the case of SARS-CoV-2, mutation rate was found to be slow, at
the level of $1.3\times10^{-3}$ substitutes per site per year\cite{Li+20b},
slower than the Influenza A/H1N1 pandemic virus which mutates at $4.4\times10^{-3}$
substitutes per site per year\cite{Su+15}. Unlike the \textquotedblleft antigenic
shift\textquotedblright{} in fast mutating viruses that thwarts cross
immunity and requires yearly vaccination\cite{Fer+03}, a slower mutation
rate may allow for the co-existence of closer strains, retaining cross-immunity,
or cross-response\cite{Nic+19,Fer+03}. We suggest that the mutation(s)
that occurred in the LVS leading to the emergence of COVID-19 affected
the pathogenicity and transmissibility of the virus, but not antigenicity.
To date, genetic sequencing has shown that SARS-CoV-2 mutated and
diverged into several genetic groups\cite{Che+20}. Mutations affecting
pathogenicity have been previously described for SARS-CoV-2\cite{Yao+20},
but currently all identified groups, to the best of our knowledge,
are highly virulent, inconsistent with the expected characteristics
of the proposed LVS. Mutations affecting transmissibility, but not
antigenicity, have been previously suggested for other respiratory
pandemics\cite{Fer+03}. Specifically, mutations effecting transmissibility
and pathogenicity were previously described for the Ebola pandemic
of 2013-2016\cite{Urb+16,Die+16}. Our proposed model therefore follows
similar pathways found earlier in other viruses.

As in those cases of cross-immunizing strains, our model suggests
that although both strains share cross immunity, the earlier strain
has mild to no clinical symptoms, leading to it being yet-unrecognized
and not raising world-wide alarm. However, when this LVS reached Wuhan,
an accumulation of mutations occurred that gave rise to the emergence
of a faster spreading and more pathogenic HVS, namely the currently
identified SARS-CoV-2, which then spread from Wuhan to China and the
rest of the world. Our model suggests, given the delay between the
onset of the LVS and the HVS, that countries with high exposure to
China could achieve earlier LVS induced herd-immunity (or near herd
immunity), leading to earlier quenching of the HVS spread. In countries
that have lower exposure to China the LVS achieves herd immunity only
later, or might not achieve herd-immunity at all before the HVS becomes
widely spread. In these countries a larger COVID-19 pandemic would
be expected. Such dynamics can explain the observed negative correlation
between traffic from China and the infection-levels in each country.
Similarly, since herd immunity is dependent on the percentage of population
that is infected by a virus\cite{Het00}, less-populated countries
could achieve herd immunity of the LVS earlier on, leading to the
positive correlation between COVID-19 infection-level and the population-size. 

As we show below, in our double-strain co-infection model, the overall
infection level for any given country is prescribed by the mobility-exposure
level of the country and its population-size, where the specific weight
of these two parameters is determined by the spread rate of the LVS.
Using a linear-regression analysis we can identify the LVS spread
rate, and explain the current infection levels for the countries in
our sample, which extends over four orders of magnitude in the number
of cases. In Fig. \ref{fig:Correlation}b we show the correlation
between the number of (testing-normalized) confirmed cases (see Methods)
and the risk-parameter, which we define below (adjusted ${\rm R^{2}=0.73};$
F-statistics vs. a constant model score of 112, with p-value of ${\rm p=}5.2\times10^{-13}$;
41 observation points, 39 degrees of freedom; using the 'fitlm' function
in Matlab\cite{MATLAB}) allowing us to explain (and predict) the
past (and future) dynamics of the pandemic. 

\begin{figure}
\includegraphics[scale=0.3]{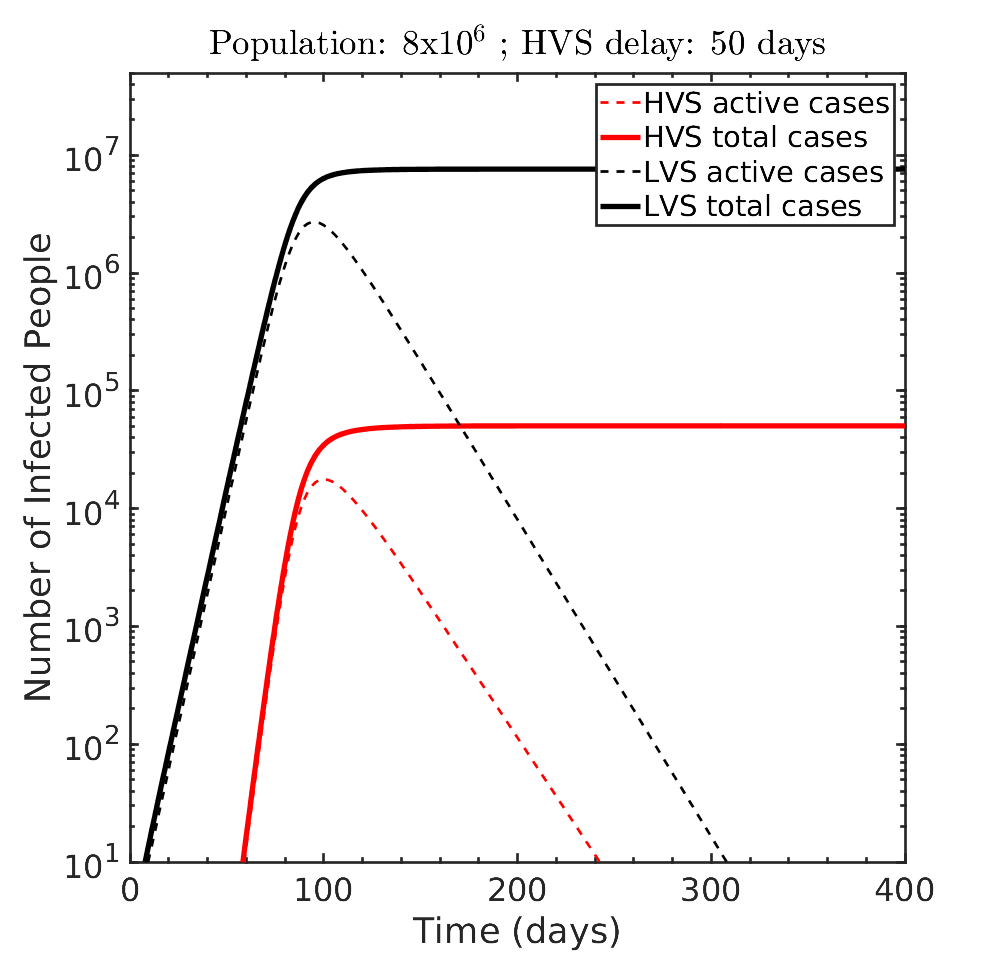}\includegraphics[scale=0.3]{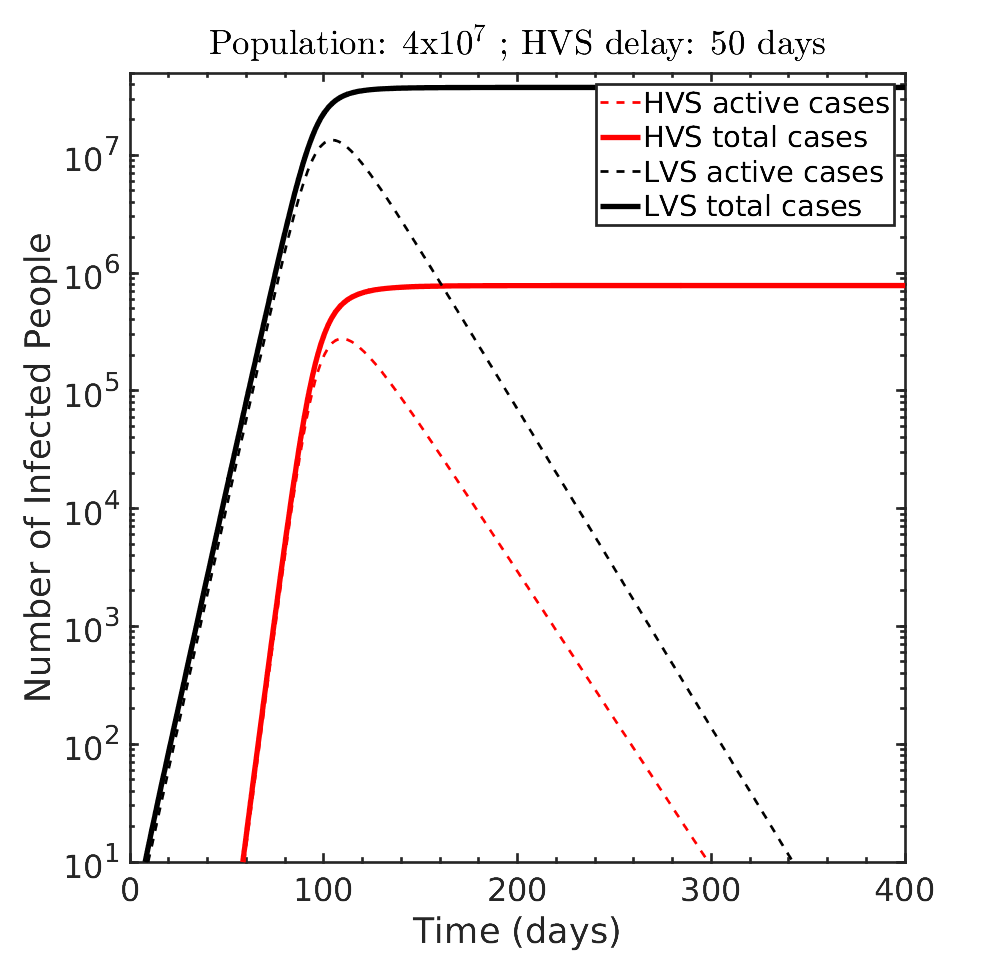}

\includegraphics[scale=0.3]{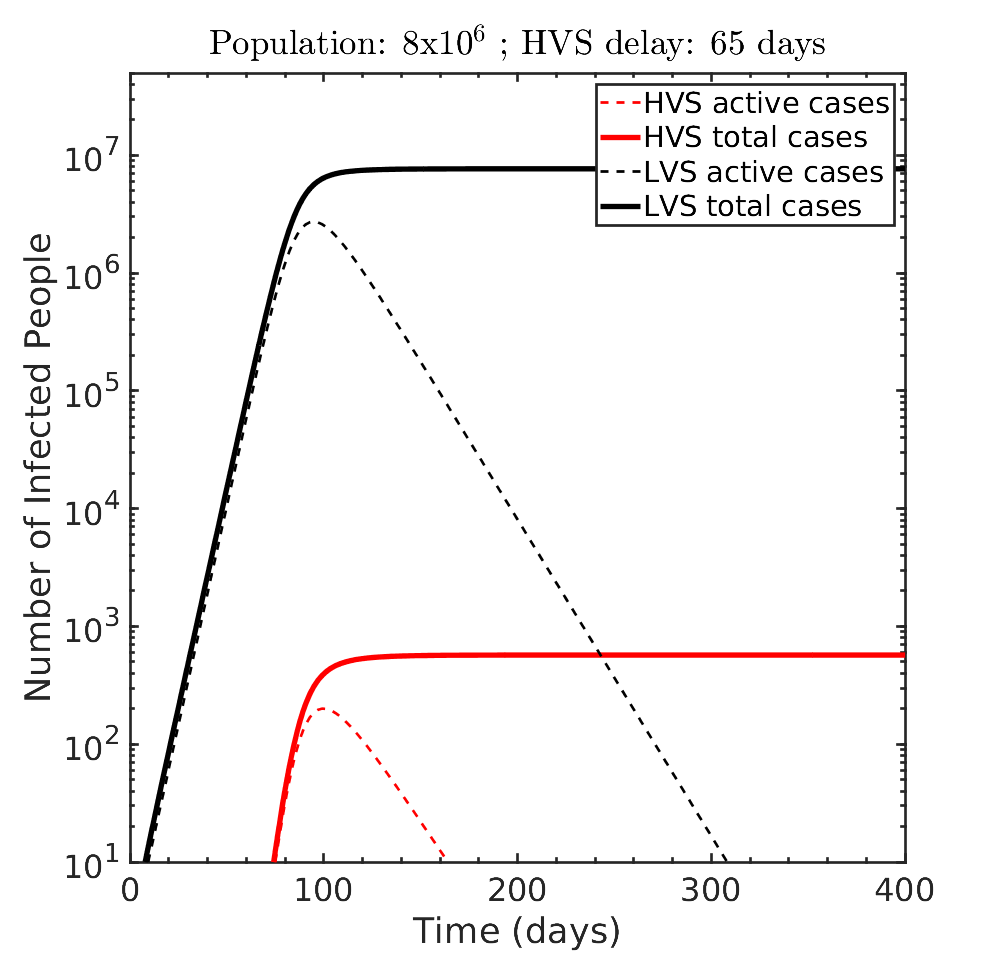}\includegraphics[scale=0.3]{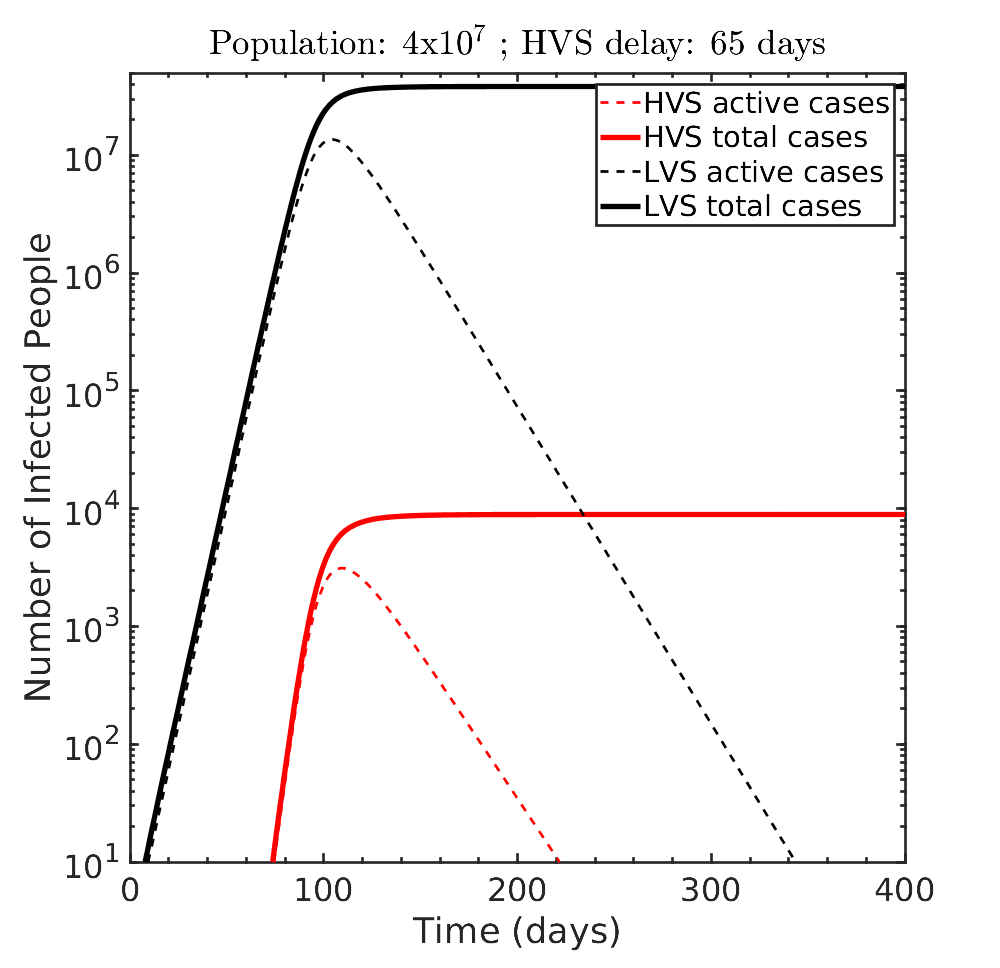}\caption{\label{fig:model}The results of a simple SIR co-infection model,
showing the evolution of two viruses which share cross-immunity, where
the slower-spreading one (the LVS, in our case), begins spreading
earlier. We show four different cases as to exemplify the dynamics.
Top (bottom) left : The LVS begins spreading 50 (65) days earlier
than the HVS in a population of $8\times10^{6}$people. Top (bottom)
right: The LVS begins spreading 50 (65) days earlier than the HVS
in a population of $4\times10^{7}$people. As can be seen, the longer
the delay between the onset of the LVS and the HVS (which is a function
of the mobility-exposure level, see Eq. \ref{eq:t_exp}), and the
smaller the population, the spread of the HVS is quenched earlier,
leading to an overall lower infection-level of the HVS (see test for
further discussion). }
\end{figure}

In order to study the dynamics of two-virus spread with cross antigenicity,
we utilized the well-established Susceptible-Infected Recovered (SIR)
model \cite{Ker+27} (see ref. \cite{Het00} for an overview). We
extended the SIR model to the case of two circulating cross-immunizing
viruses (i.e. a person infected by one of the viruses is immune to
infection from the other virus, both during and after the infection;
see Methods and related previous studies of co-infection and the spread
of multiple strains\cite{Die+79,Cas+89,Adl+91,Bal+11,Sus+15,Kuc+16,Nic+16,Tho+19,Chi+20}).
In the model we track the number of people susceptible to both viruses
(S), the numbers of people infected with the LVS, or with the later-emerging
HVS, (${\rm I_{L}}$ and ${\rm I_{H}}$, respectively), and the number
of people who have had contracted either strain, and then recovered
or died (${\rm R_{{\rm L}}}$ and ${\rm R}_{{\rm {\rm H}}}$, for
the LVS and HVS, respectively). It is assumed that the total population
in a given country $N=S(t)+I_{L}(t)+I_{H}(t)+R_{L}(t)+R_{H}(t)$ is
fixed, and that those who have recovered from either virus are immune
to both. As we discuss later on, we assume the precursor LVS spreads
slower than the HVS. The probability of a person contracting a given
virus at time $t$ is proportional to the fraction of susceptible
people among the total population ${\rm I_{H}/N}$ for the HVS, and
${\rm I_{L}/{\rm N}}$for the LVS. These assumptions lead us to a
set of five ordinary differential equations for $S(t)$, $I(t)$,
and $R(t)$: 

\begin{eqnarray}
\frac{dS}{dt} & = & -\beta_{L}\frac{S(t)I_{L}(t)}{{\rm N}}-\beta_{L}\frac{S(t)I_{L}(t)}{{\rm N}}\label{eq:2a}\\
\frac{dI_{L}}{dt} & = & \beta_{L}\frac{S(t)I_{L}(t)}{{\rm N}}-\gamma_{L}I_{L}(t)\label{eq:2b}\\
\frac{dI_{H}}{dt} & = & \beta_{L}\frac{S(t)I_{L}(t)}{{\rm N}}-\gamma_{H}I_{H}(t)\label{eq:2c}\\
\frac{dR_{L}}{dt} & = & \gamma_{L}I_{L}(t)\label{eq:2d}\\
\frac{dR_{H}}{dt} & = & \gamma_{H}I_{H}(t)\label{eq:2e}
\end{eqnarray}
Here, $\gamma_{L},\,\gamma_{H}\geq0$ are the recovery/death rates
and $\beta_{L},\beta_{H}\geq0$, the transmissibility parameters,
measure the likelihood of transmitting the LVS and HVS, respectively,
when an infected and a susceptible person come into contact. Here
we make a simplifying assumption that a person can transmit the virus
to other people for a total period of 14 days, before the person recovers
(or dies). While the LVS has low/negligible-virulence, as discussed
earlier, and is assumed, for simplicity that infected people recover
at the same rate from both viruses. In both cases we assume for simplicity
that no measures to slow or stop the spread of either virus have been
implemented. As can be seen in Fig. \ref{fig:Correlation}b, our results
provide an excellent fit for the data, even without considering any
containment measures, suggesting that the latter might have only a
relatively small effect on the final infection level outcome, although
they could give rise to an overall slower progression of the pandemic
(see Methods for further discussion of our assumptions). Examples
for the evolution of such double-strain co-infection model dynamics
are shown in Fig. \ref{fig:model}.

As can be seen in Fig. \ref{fig:model} (see also methods for further
details), the larger the initial spread of the LVS (LVS seed-infection)
at the time when the HVS began spreading in a given country, the faster
the LVS achieves herd-immunity, thereby also quenching the fast spread
of the HVS, and giving rise to a lower HVS-infection level. Since
herd-immunity requires a population-wise infection-level (e.g. $\sim1-1/\beta\gamma$
of the population; see \cite[ and references therein]{Bri+20}), more
populated countries take longer to achieve herd-immunity for the same
seed-infection (Fig. \ref{fig:model}c \& Fig. \ref{fig:model}d).
This allows for more time for the spread of the HVS, giving rise to
an overall higher HVS infection-level. Since the spread of the HVS
is faster than that of the LVS, the former (HVS) can even outrun the
latter (LVS), in which case the HVS will not be limited by the LVS,
and will only be limited by the population size (See Fig. \ref{fig:model}c).

Based on the double-strain co-infection model we can now explain the
observed distribution of COVID-19. The LVS outbreak began in China,
far from Wuhan. Shortly after the LVS reached Wuhan, and before the
LVS achieved herd immunity in this region (e.g. similar to the case
in \ref{fig:model}b), cumulative mutations lead to the emergence
of the HVS in Wuhan, initiating the initial exponential spread of
the HVS strain and its outbreak in China and later worldwide. At later
time the HVS spread slowed down and almost completely stopped once
the LVS achieved herd-immunity, quenching the exponential spread of
both the LVS and the HVS. Note that in the model the total \emph{cumulative}
number of infections can still slowly rise by additional factor of
2-3 after herd-immunity is achieved (which approximately corresponds
to the time when the peak in active-cases is seen; see Fig. \ref{fig:model}).
Throughout China the LVS had more time to spread before the spread
of the HVS, leading to immunization of an even larger fraction of
the population (compared to Wuhan and Hubei region) before the spread
of the HVS from Wuhan, leading to an even lower COVID-19 infection
rate in China outside Hubei province. 

The global spread of the LVS followed the exposure level, with more
exposed countries receiving LVS-infected people earlier (strong negative
correlation between the logarithm of the exposure level and the day
of first infection (see methods and Fig. \ref{fig:Exposure-vs-first-infection};${\rm {\rm p}=6.5\times10^{-10}}$).
If the LVS and HVS spread rates were the same, the spread of both
strains would have followed the exact same dynamics, and the delay
between the arrival of the LVS and the HVS would be constant. Consequently,
a fixed ratio of LVS-infection level to HVS-infection level in different
countries (with similar population sizes - see below) would be expected,
leading to an overall similar HVS-infection levels. However, the infection
levels in countries of similar population sizes differs between lower
and higher exposures (e.g. compare France and Thailand, Taiwan and
the Netherlands in Fig. \ref{fig:Correlation}). Therefore, the observed
differential infection level, requires the LVS transmissibility level
to be lower (consistent with our finding below). In this case, as
shown In Fig. \ref{fig:model} and discussed in the Methods, the HVS
spread can outrun the LVS spread and decrease the relative LVS to
HVS infection levels with time. The LVS spread to less exposed countries
started later on, allowing for a shorter time frame for the faster
HVS spread to outrun the LVS spread, giving rise to a differentiated
increase in the ratio of HVS-to-LVS infection with decreasing exposure-level.
For a given population size this ratio determines the final HVS infection
level (compare Figs. \ref{fig:model}a and \ref{fig:model}c). 

Another important parameter that determines viral spread in the double-strains
co-infection model, is population size. As discussed above, the time
to achieve herd immunity depends logarithmically on the population
size. Therefore, countries with smaller population size would approach
herd immunity faster, allowing less time for the HVS spread to outrun
the LVS-spread before its exponential growth is quenched. 

Considering the exposure-level - dependent HVS to LVS infection ratio,
the population size, and the initial delay between the onset of the
LVS and the onset of the HVS in China, we expect the HVS infection
level to have a specific dependence on these parameters (as we derive
in details the Methods), given by 

\begin{equation}
\log_{2}({\rm I}_{{\rm H}}^{{\rm max}})=[(t_{0}^{L}-t_{0}^{H})/{\rm T}_{{\rm H}}]+\left(\frac{{\rm T_{L}}}{{\rm T_{H}}}\right)\log_{2}{\rm N}-\left(\frac{{\rm T_{L}}}{{\rm T_{H}}}-1\right)\log_{2}{\rm c}_{exp},\label{eq:IHmax-log-1}
\end{equation}
where ${\rm I}_{{\rm H}}^{{\rm max}}$ is the expected maximal infection
level of the HVS, $t_{0}^{L}$ and $t_{0}^{H}$ are the onset times
of the LVS and the HVS, respectively; and ${\rm T}_{{\rm L}}$ and
${\rm T}_{{\rm H}}$ are the doubling times for the LVS and HVS, respectively.
In particular, we expect a specific relation between the dependence
on the population size, and on the doubling times ratio (${\rm T_{L}}/{\rm T_{H}}$compared
with ${\rm T_{L}}/{\rm T_{H}}-1$; see Methods). Linear regression
can therefore allow us to identify the specific dependence. Using
the data on confirmed cases we find best fit linear regression models
(for Eq. \ref{eq:IHmax-log-1}) of ${\rm T_{L}/T_{H}=}1.31\pm0.23$
(Adjusted ${\rm R}^{2}$ of 0.42). Although already providing highly
significant results (${\rm p}=5.3\times10^{-9}$; Pearson correlation
test), such data are noisy due to differences in the testing level.
When analyzing the testing-normalized confirmed cases (see methods)
we find best fit linear regression models (for Eq. \ref{eq:IHmax-log-1})
of ${\rm T_{L}/T_{H}=}1.59\pm0.17$ (Adjusted ${\rm R}^{2}$ of 0.74).
Using the age-normalized deaths data we find ${\rm T_{L}/T_{H}=}1.57\pm0.32$
(Adjusted ${\rm R}^{2}$ of 0.41), or, when also age \emph{and} testing-normalized
data, giving ${\rm T_{L}/T_{H}=}1.72\pm0.26$ (Adjusted ${\rm R}^{2}$
of 0.62). All of these are consistent with each other and the model.
Hereafter, we adapt the most significant result, ${\rm T_{L}/T_{H}=1.59}$,
as our fiducial ratio. Correlation tests between the risk-parameter,
$\chi$, which we define (following Eq. \ref{eq:IHmax-log-1}) as
\[
\chi=\left(\frac{{\rm N}}{{\rm 2^{(t_{0}^{H}-t_{0}^{L})/T_{L}}c_{exp}{}^{1-T_{H}/T_{L}}}}\right)^{{\rm T_{L}/T_{H}}},
\]
and the number of testing-normalized cases (age and testing- normalized
deaths) for ${\rm T_{L}/T_{H}=1.5}9$ give ${\rm {\rm p}=5.2\times10^{-13}},$
$3.8\times10^{-10}$ and ${\rm {\rm p\ll1e-20}}$ (${\rm p=1.8\times10^{-10},\,1\times10^{-7}}$
and ${\rm p=}1.4\times10^{-7}$), for the Pearson, Kendall, and Spearman
tests, respectively. 

Taking the mobility exposure-level in China to be unity, by definition,
and plugging the population size and the infection level, we can then
find the delay time between the onset of the LVS and the HVS. If we
assume a doubling time of $2.7$ days (in the absence of any containment
measures), consistent with observed spread rate at the earliest days,
we find $(t_{0}^{L}-t_{0}^{H})\sim86$ days, suggesting the initial
human infection by the LVS began in China around September. This,
however does not account for non-diagnosed infections, population
structure and inter-China mobility that may slow transmission, and
it is possible that the delay might somewhat differ. If we assume
only $\sim0.1$ of of the infections are diagnosed we get $\sim77$
days. 

In Figs. \ref{fig:Correlation}c and \ref{fig:Correlation}d we show
${\rm I_{H}^{max}}$ for both the testing-normalized confirmed cases
and for the age and testing-normalized deaths, as a function of the
risk-parameter defined above. In fact, low and high population countries,
not used to derive the model parameters as not to introduce potential
bias (see Methods) also well fit (including them in the statistical
score gives ${\rm p=7\times10^{-22}}$ (Pearson) for the same model
parameters), and the model shows an excellent agreement with observations,
over five orders of magnitudes in infection level. We find that that
in most countries in our sample the spread of the LVS should achieve
herd-immunity level, and quench further exponential spread of the
HVS, much before the HVS achieves herd-immunity infection-level. We
find low dispersion around the expected values from the model (up
to a factor of a few for the testing-corrected confirmed cases), leaving
relatively little room for other parameters beside exposure level
and population size in affecting the final outcome, as we further
discuss below. A wider dispersion can be seen in the age-corrected
deaths. However, the amplitude of the dispersion grows with the number
of deaths/confirmed-cases, suggesting a relation to the load on the
health system affecting treatment, and thereby the number of deaths.
A further inquiry on this issue, though important, is beyond the scope
of this paper and will be explored elsewhere.

Our model and analysis of the current statistical data supports the
existence of a preceding LVS. They explain and predict upper limit
for infection-levels as a function of the mobility-exposure to China
and the population-size. In particular, we find that more exposed
countries the spread of the LVS is already at or close-to herd-immunity
level, which explains the puzzling observed low-level infections in
these countries, and the direct, but non-linear dependence of infection
level on population size, which is otherwise unexpected given the
low infection level in respect the those required for herd-immunity.
However, additional independent tests could further support or refute
the model. These can be divided between biological tests and demographic
one, as we discuss in the following. 

Cross antigenicity between the LVS and COVID-19 is an essential part
of our model and can be either humoral or cellular. If cross antigenicity
relies on cross reactive antibodies, it is possible that antibodies
to LVS will be detected by serological testing for SARS-CoV-2, already
employed to some extent in several countries\cite{Ioa+20}. In particular
countries which appear to have achieved LVS herd-immunity should have
shown a large fraction of the population to be sero-positive, in contrast
with the currently directly measured lower per-population HVS infection-levels\cite{Ioa+20,Hav+20}.
However, current serologic testing is optimized for HVS, and might
be less efficient for detection of the LVS. Recent findings at the
levels of a few percents up to 30 percents sero-positive, i.e. tens
to hundred times larger than the infection rates inferred from the
confirmed cases, but still considerably lower than required for herd-immunity.
These likely reflect low-testing levels in most countries, suggesting
that actual HVS-infection levels are typically much higher than inferred
from the reported cases, consistent with our use of testing-normalization
in our analysis (see Methods). It also suggests that none of the currently
identified genetic groups of SARS-CoV-2 could be the LVS proposed
here. Moreover, it is important to note, that it is currently unclear
whether the antibodies detected in the currently available serologic
tests are indeed protective antibodies, and therefore even in the
cases of humoral cross antigenicity, LVS strains might not be detected
at all by SARS-CoV-2 serologic tests. We therefore suggest to make
use of a more direct and accurate method to test the existence of
LVS antibodies, by employing a viral micro-neutralization testing
of the HVS on serum from a sample of people (preferably from highly
exposed countries, for which the majority of the population should
already have been infected by the LVS) who are found to be negative
for the SARS-CoV-2 HVS in serologic tests. These should be able to
identify antibodies reaction to the HVS otherwise undetected by currently
employed serologic test, in a similar manner as used to test acquired
immunity to a high pathogenic flu virus following vaccination for
a less pathogenic flu virus\cite{Ste+05}. To the best of our knowledge
only one micro-neutralization study (with limited statistics) has
been done on SARS-CoV-2 to date\cite{Man+20}, not identifying antibodies
in the serum of people who were found to be negative for the SARS-CoV-2
HVS in serologic tests.

A likely possibility is therefore that cross antigenicity between
the two strains might be dependent on \emph{cellular} immunity, rather
than humoral one. For example, previous studies on avian influenza
HVS H5N1 protection by the LVSs H9N2\cite{Seo+01,Kha+09}, H1N1, or
H1N2\cite{Nfo+12} showed that the immunity was cellular rather than
humoral immunity, and H5N1 antibodies were not found in chickens that
were previously infected by low virulence strains\cite{Seo+01,Kha+09,Nfo+12}.
In fact, very recently, and after the completion of our analysis Sekine
et al. \cite{Sek+20} detected SARS-CoV-2-specific T cells were detectable
in antibody-seronegative family members and individuals with a history
of asymptomatic or mild COVID-19. This could be consistent with our
suggestion of cellular cross-immunity and the possible wide-spread
immunity due to the LVS. Our model suggests that mapping T-cell response
test among a current sample of people with no known infection in most
countries and especially in highly exposed countries (e.g. China,
Vietnam, Thailand) should show the majority to already be immune,
inconsistent with the known low-level infection level in these countries
in respect to the population size, verifying our results. Although,
at this stage, one can not exclude acquired immunity through past
infections from other viruses, however, such immunity level should
not follow the specific correlations expected from our model and explaining
the current global distribution. 

Assuming a LVS SARS-CoV-2 is identified and a specific serologic or
mapping T-cell response test is developed for it, tests of blood samples
taken across China (or in other countries achieving LVS herd-immunity)
around October-November, i.e. before the December 2019 HVS outbreak,
but at the point where the LVS should have already been widely spread,
could identify positive cases, in particular outside Hubei province.
In fact, the study of samples taken at different times would potentially
allow for identifying the LVS spread, where positive fractions should
gradually increase from September until November. Similarly, early
evidence for SARS-CoV-2-like viruses found before the known COVID-19
outbreak in December 2019 could further support our finding, e.g.
through finding evidence for SARS-CoV-2 in the sewage, much before
the COVID-19 outbreak began at the respective country\cite{Fon+20},
could provide similar clues of the early spread of the LVS. In particular,
a much earlier spread of the HVS should have been easily identified
due to the large number of patients expected, inconsistent with the
data, while a wide spread LVS, could be consistent with such findings. 

Biological tests could provide optimal smoking-gun signatures of the
LVS spread, but are depend on the unknown genetic similarity of the
LVS and the HVS, the type of serologic or mapping T-cell response
tests and the type of cross antigenicity. A different, sequence-independent
statistical test approach can be used to test for LVS spread. People
visiting China in the 1-2 months before the HVS outbreak (but not
afterwards), might have been infected by the LVS during their visit,
thereby acquiring immunity to the HVS. HVS infection-level among such
people should therefore be lower than that of the background population
in their home country, in particular in COVID-19 highly infected regions.
Therefore testing COVID-19 prevalence among people visiting China
in September to November 2019 could confirm the presence of a protective
LVS strain independent of its genomic sequence, but requires non-trivial
data collection due to the large travel and health information needed
for this statistical analysis. 

The existence of a LVS providing cross-immunity to the COVID-19 HVS
strains has far-reaching implications. First, it proves the existence
of acquired viral immunity and the potential for developing an efficient
vaccine, given the immunity provided by the LVS as an effective live-vaccine.
Biological identification of the LVS could be used as a highly advantageous
starting-point for developing a safe COVID-19 vaccine, although modifications
preventing the occurrence of new pathogenic properties of the natural
virus will be essential. Moreover, although the LVS might protect
from further infections in already ``protected'' countries, it is
yet unknown how long such immunity protection lasts (see e.g. discussion
of limited immunity in Ref. \cite{Hua+20}). Lastly, the HVS SARS-CoV-2
might at some point change its structure to generate a novel HVS that
would not be cross immune with the previous strains.

Biological identification of the LVS would also enable the development
of specific serologic tests and/or mapping T-cell response tests,
that can then be used to measure and directly validate the existence
of herd-immunity in a given country, and thereby direct the application
of containment measures and opening strategy plans. 

Non-homogeneous geographical transmission and/or non-homogeneous transmission
among less/disconnected communities can give rise to pockets of less
immune populations within largely-immune countries, explaining the
possibility of local outbreaks of the HVS. More generally, our simplified
model considers countries as independent units, while intra-country
dynamics could lead to a down-scaled version of the model. Even in
countries in which the early LVS infection provided partial protection,
the protection might not be homogeneous. Regions and/or communities
more exposed to international traffic would be infected earlier, leading
to initially higher infection levels, but eventually show lower infection-level
at later times, since the same regions/communities also acquired higher
level of immunity through larger prior LVS infection. More isolated
regions and/or remote/disconnected communities might eventually be
at higher risk and could have higher infection levels compared with
the overall country population. Generally, regions/communities having
earlier infections are likely to experience overall lesser cumulative
levels of infections, compared to same-size communities infected later,
and larger isolated communities are generally likely to experience
higher infection level. Similar intra-country analysis (e.g. comparing
the exposure levels and population-sizes in different states in the
US, would suggest some of the more populated states in the US are
still likely to experience higher infection levels), could serve to
predict the intra-country dynamics of the COVID-19 pandemic. Furthermore,
COVID-19 can still affect immune deficient people everywhere, as these
are not protected by the LVS, although herd-immunity significantly
lowers their chance of infection.

It is possible, and even likely that in a fraction of the cases the
LVS only provides partial immunity, allowing for HVS infection, but
leading to a less virulent form of COVID-19. In such a case we might
expect a higher fraction of newly identified COVID-19 patients to
be, on average, more symptomatic during the early phases of the pandemics
when it still spreads exponentially, before the LVS achieves her-immunity.
At these early times the LVS has not yet infected the majority of
the population and newly HVS-infected people are not likely to have
been previously infected by the LVS, and be partially immunized. After
the LVS infected a large fraction of the population, new HVS-infections
are far more likely to be of previously LVS-infected people, who already
acquired partial immunity. We might therefore expect a lower fraction
of asymptomatic cases, and a higher morbidity rate during the early
exponential growth of the HVS, in comparison with later times after
sub-exponential-growth and later decay in the number of cases is observed.
We do note that the improvement in COVID-19 treatment in time due
to the learning curve of the health systems might give rise to lower
morbidity at later times too, but it is less likely to to affect the
asymptomatic to symptomatic ratio, nor should it be related to the
transition from exponential to sub-exponential growth. Note, that
in a case of only partial immunity, the infection-level, in terms
of cases, could somewhat increase beyond the level described above,
the transition to sub-exponential growth, but the infection-level
in terms of deaths is likely to be far less affected. 

Our results indicate that the infection level depends on the exposure
level and the population size. In particular there is relatively little
dispersion in the observed infection levels in respect to the basic
model fit. This suggests that although the widely varied containment
measures applied by different countries\cite{Hal+20} might be important
in slowing the virus spread to some extent, and possibly allowing
the health system to better accommodate more serious cases, the specific
and different measures taken by each country eventually had little
effect on the number of infected people in a given country. In particular,
it is difficult to understand how would the different measures taken
at different times could give rise to the type of correlations with
exposure-level and population sizes which we identify. Our results
would similarly suggest that different opening strategies after lock-downs
and social distancing would not considerably affect the overall infection
levels, which is effectively dependent on the double-strain dynamics
and not the containment measures responses (or the lack of them).
While it is not clear whether any of the currently used containment
measure had a significant effect on the overall number of infected
people, it is important to note that in principle, measures which
could be more effective for eradicating the HVS compared with the
LVS should be prioritized. Such measures may change the relative LVS
to HVS infection levels, and would provide the same effect as introducing
a higher per-population exposure, and in turn constrain the maximal
HVS infection level to lower values. Current contact tracing, for
example, is focused on HVS infections, while not affecting the LVS
spread; similarly, quarantine/isolation of HVS-infected people decreases
the HVS-transmittibility. Social distancing, however, affects both
the LVS and the HVS, and might even be harmful under some conditions,
as it lowers the transmissibility parameter $\beta$ for both the
LVS and the HVS. Given that the doubling time and $\beta$ are related
through ${\rm T=log(2)/log(1+\beta)}$, one gets ${\rm T_{L}/T_{H}=log(1+\beta_{H})/log(1+\beta_{L})}$.
Decreasing both $\beta_{{\rm H}}$ and $\beta_{L}$ by the same factor
increases ${\rm T}_{{\rm L}}/{\rm T}_{{\rm H}}$ allowing the HVS
to outrun the LVS faster, i.e. leading to an overall higher HVS infection-levels.
However, this is only true as long as the reproduction number is bigger
than unity and growth is exponential, if the containment measures
are sufficient to drive the reproduction number below one, the exponential
spread of the disease is stopped in any case, and than these measures
are indeed helpful. 

Fast mutating viruses might mutate too fast as to give rise to long-lasting
cross-immune strains. The slow mutation rate of SARS-CoV-2 could be
the reason allowing for LVS and HVS co-infection. Nevertheless, any
new low-virulence viruses are likely to go through several mutations
before becoming virulent. This would suggest that continuous monitoring
of viruses mutations in the population could be used both to identify
the potential progress into virulent forms, as well as serve for development
of new vaccine, once a virulent virus forms. In particular, sample
collections of non-virulent viruses and their sequencing done continuously
could serve as key input in the onset of a new epidemic, allowing
for immediate back-tracking of the new-virus development, and the
possible use of earlier non-virulent strains as an advanced starting
point for development of a vaccine. 

Finally, although developed in the context of the COVID-19 pandemic,
similar multi-strain analysis and the statistical identification of
co-infection described here can be used for the study of other pandemics.

\section{Methods}

\subsection{Data sources}

\label{subsec:dat}

We have used several data sources for this analysis, described herein
. 

COVID-19 cases and tests and population sizes: The numbers of confirmed
COVID-19 cases, deaths, and number of tests, as a function of time,
as well as the data for population sizes were taken from the our world
in data (OWID) website (www.ourworldindata.org), which (COVID-19)
data is based on the EU-CDC data (https://www.ecdc.europa.eu/en),
which we downloaded on 4/7/2020. For verification, we also compared
the data from this site to that in the worldometer website (https://www.worldometers.info/coronavirus/)
and find it consistent with the data we used, besides the number of
tests for Peru, which is much higher in the worldometer website. For
consistency, in our analysis shown here we only make use of the OWID
data, which also provide the full data as a function of time since
the beginning of the pandemic, and not only the last few days as done
in the worldometer data. Nevertheless, we did make the same analysis
on the worldometer data for verification (the data also contain testing
data on several countries not available in the OWID data), and found
fully consistent results, i.e. the same doubling time ratio, and at
a similar level of significance, further verifying the data reliability
and consistency.

Mobility-exposure level: The main analysis is based on exposure data
derived by the open tool web-platform EpiRisk module at https://epirisk.net/
taken on 3/6/2020. As described on the web-site, EpiRisk is a computational
platform designed to allow a quick estimate of the probability of
exporting infected individuals from sites affected by a disease outbreak
to other areas in the world through the airline transportation network.
EpiRisk is part of the the Global Epidemic and Mobility project (GLEAM;
https://www.gleamproject.org/) which combines real-world data on populations
and human mobility as to inform epidemic models and was already used
for studying COVID-19 spread\cite{Chi+20}. These data provided us
with exposure level estimates for over 100 countries with the highest
air-transport mobility rates incoming from China. EpiRisk mobility
data also accounts for seasonal differences, and we considered both
risk model in the Autumn (August-October) and Winter (November-January),
generally providing very similar results; the results shown in the
paper are for the Winter mobility results, at which time most of the
infections were likely to occur.

In order to enable an easy reproduction of our results we provide
the EpiRisk mobility-exposure data (for the Winter season). We also
provide the risk-parameter we derive for each country in the same
EpiRisk mobility-exposure file. In addition we provide the data on
the number of COVID-19 cases, deaths, population-size, number of tests
and age-structure from the ``Our world in data'' website, taken
on July 4th 2020. 

\subsection{Data filtering}

In our analysis we considered data only for countries for which our
above mentioned data sources provided all relevant information, i.e.
number of confirmed cases, number of deaths, number of tests and mobility-exposure
levels, and population sizes (the latter was available for all countries).
We then made use of several filtering and normalization methods to
minimize data noise level , as described below. 

\subsubsection{Age and testing normalization }

The exact numbers of COVID-19 infections are unknown, since screening
of the entire population has not been done in any country. We therefore
use the numbers of confirmed cases and the numbers of COVID-19 related
deaths as proxies for infection levels. However, each of these numbers
is influenced by various factors, which are difficult to asses, and
their exact relation to the actual infection level could differ to
some extent in different countries, giving rise to effective noisy
data. We made of two corrections in order to filter some of the main
components. 

COVID-19 related deaths are typically of patients which were earlier
critical condition. These are more likely to have been identified
and tested than asymptomatic patients, which might have never been
tested. Confirmed deaths are therefore potentially more reliable proxy
for the infection level. On the other hand the majority of the deaths
are of people older than 70, and therefore differences in age structure
in different countries will give rise to different ratio between the
infection level and the number of deaths. We therefore age-calibrated
the death rates in each country by dividing the number of deaths by
the fraction of the population older than 70 years. Other potentially
important affecting factor is the quality of the health system, where
better health systems might be able to provide better treatment and
decrease the morbidity level. Finally, even in high quality health
systems, an overload of a large number of patients in critical conditions
might affect the quality of treatment, potentially leading to higher
morbidity rates in various countries. The two latter factors are more
difficult to asses quantitatively, and are not accounted for in our
analysis. We only considered countries for which our data source provided
the age-structure.

\begin{figure}
\includegraphics[scale=0.3]{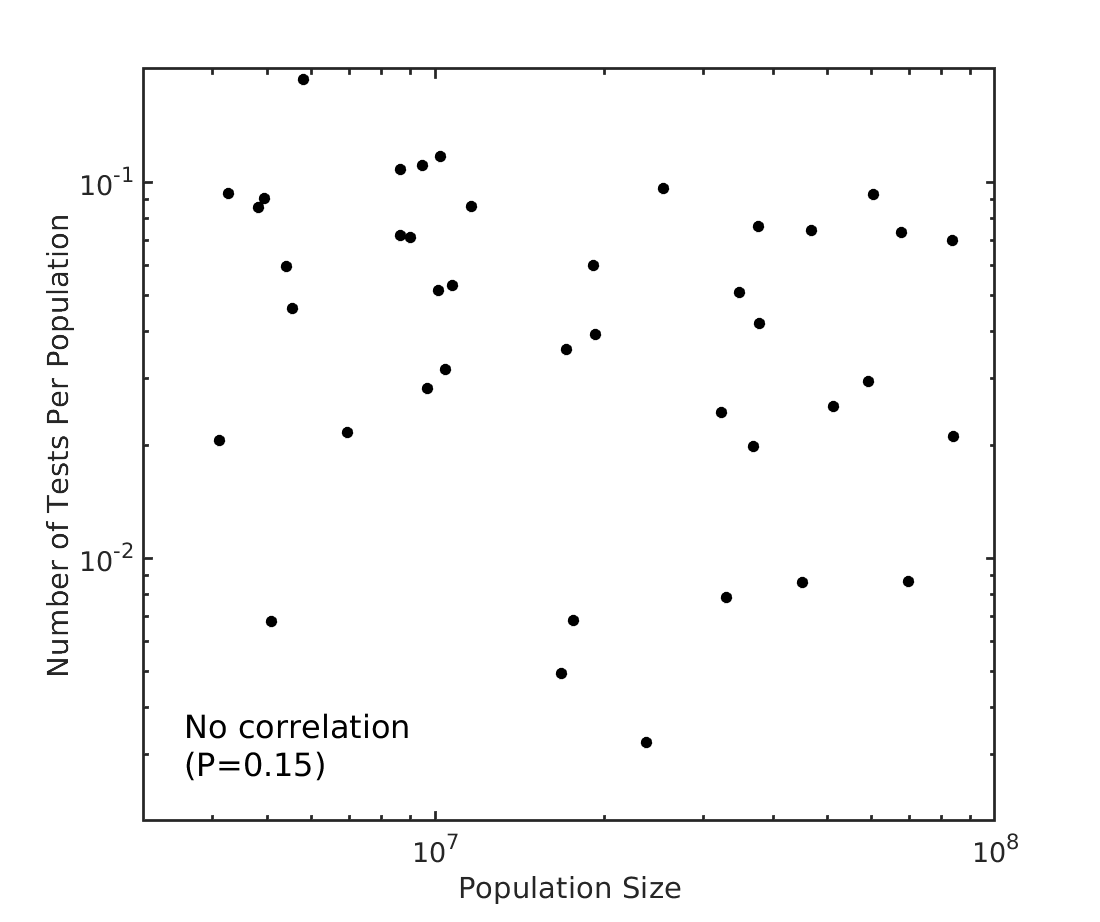}

\caption{\label{fig:Testing-per-population}The relation between population
size and the number of COVID-19 tests per population. As can be seen
we find no correlation (${\rm P=0.1}$) between these parameters.}
\end{figure}

The number of confirmed cases is likely to be less affected by the
overall quality of the health system, but is more dependent on the
number of COVID-19 tests done in respect to the population size, and
the testing strategy. We therefore normalized the numbers of confirmed
cases in each country by dividing them by the corresponding numbers
of tests per population in these countries. Since our model predicts
a relation between infection level and the population size, one might
be concerned that inclusion of such a parameter which is related to
population size could introduce an artificial correlation. We find
that the tests per population show no correlation with the population
size (${\rm p}=0.15$; see also Fig. \ref{fig:Testing-per-population})
when excluding the most ($>10^{8}$ people) and least ($<2\times10^{6}$
people) populated countries in our sample. In Fig. \ref{fig:Outliers}
we show the data for all countries in our sample, including these
high/low population ones. As can be seen, these data for these countries
is highly consistent with the model. In fact, when adding these countries
to the analysis (but not the systematic outliers (see below) also
shown in the figure) the results even further improve, and we find
adjusted-$R^{2}=0.84$ and Pearson correlation test p-values of $4\times10^{-22}$,
for the same model parameters derived without considering these countries. 

\subsubsection{Mobility-exposure level}

\begin{figure}
\includegraphics[scale=0.3]{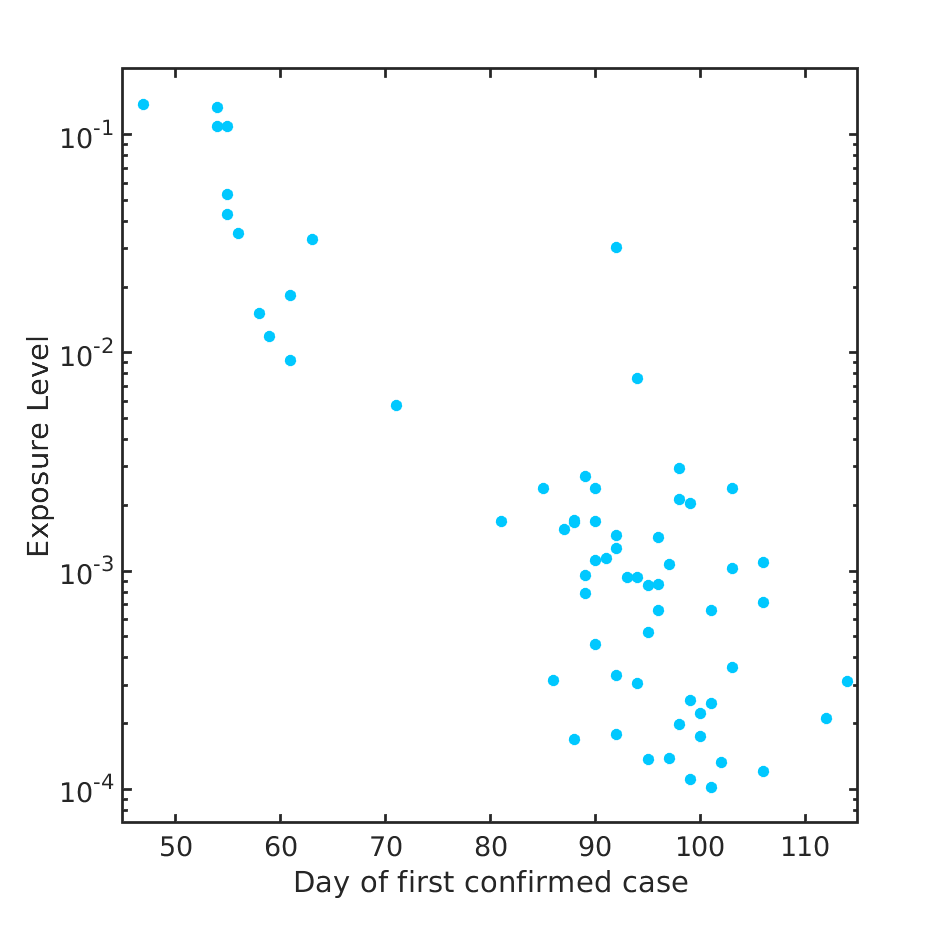}

\caption{\label{fig:Exposure-vs-first-infection}The day of first confirmed
infection for each country as a function of the exposure level to
China. }
\end{figure}

Throughout our analysis we assume that each country is exposed to
the viruses directly through traffic from China. While this is likely
to true for countries highly exposed to China, less exposed countries
might be infected via secondary spread from other countries, even
before being infected through direct incoming traffic from China.
A detailed pandemic model which accounts for the traffic between each
country and the epidemic evolution in time could potentially better
account for such secondary exposed countries. This is beyond the scope
of the current study, and we therefore only account for countries
which were likely to be directly exposed to China. In Fig. \ref{fig:Exposure-vs-first-infection}
we plot the day of first confirmed infection for each country as a
function of the exposure level to China. From Eq. \ref{eq:t_exp}
(see below) we expect a correlation between the logarithm of the exposure
level and the initial infection in a given country. An overall strong
correlation can indeed be found for the full data sample (${\rm p<10^{-17}})$,
but above $\sim100$ days the correlation saturates, possibly suggesting
that countries exposed to COVID-19 more than $\sim100$ days after
December 1st 2020, are more likely to have been exposed indirectly.
Therefore, in our analysis we only make use of data from countries
in which the first confirmed case has been identified at most 100
days after December 1$^{{\rm st}}$ 2019. As can be seen a large dispersion
can already be seen before $100$ days suggesting that the exact identification
of countries with experience direct or secondary exposure is difficult
to discern without a more detailed model. Nevertheless, we find that
even making use of the full data, i.e. even beyond 100 days, does
not affect our conclusions, i.e. we can still identify (with significant
statistical score, albeit less significant than those discussed above)
model parameters consistent with those we find using the filtered
data. In principle the initial infection day might be used (after
appropriate analysis) as a phenomenological proxy for the exposure
level, irrespective of the traffic data. This is beyond the scope
of the current analysis and will explored in later studies. 

\subsection{Systematics and outliers}

\label{subsec: systematics}

Throughout our analysis we have made use of all data available to
us, as described in section \ref{subsec:dat}, and the statistical
significance levels are based on analysis of these data. Nevertheless,
as mentioned in the main text, some clear systematic effects and outliers
can be identified in respect to specific countries. It is therefore
important to discuss these systematics and the possible origin of
outliers.

Our analysis makes use of exposure-level assessments based on air-traffic
mobility from China. This may give rise to two systematic types of
incorrect estimates of the exposure level. Underestimates for China
neighboring countries, and overestimates of flight-transportation
hubs. 

China has long ground borders with neighboring countries allowing
for considerable ground traffic mobility, which data are not available
to us. Passenger trains operate from China to Vietnam, Mongolia, Kazakhstan,
Russia and North Korea. Besides trains, border crossing points allow
for ground mobility to the other neighboring countries, including
Bhutan, Kyrgyzstan, Kazakhstan, Laos, Nepal, Tajikistan, Myanmar,
India, Pakistan and Afghanistan. The exposure level of these countries
are therefore likely to be underestimated. Increasing their exposure
level could drive them to the regime of higher per-population exposure
level, and constrain and lower their HVS infection level due to the
LVS herd-immunity limit. We therefore did not include these countries
in our main analysis. 

Air-traffic data might not well account for flight-hubs, in which
significant fraction of incoming passengers are commuting through
on their way to other countries. In such cases the exposure level
might be overestimated. In more populated countries, the addition
of the commuting might not considerably affect the overall incoming
traffic level. The only cases in which the annual traffic through
the airport is at least approximately ten times larger than the population
in the host country itself are Dubai airport in the United Arab Emirates,
Singapore airport and Doha airport in Qatar (all three by a large
margin from other major hubs; according to the Airport Council International
world airport traffic and rankings). The exposure level of these flight-hubs
are therefore likely to be overestimated. We therefore did not include
these countries in our analysis. 

\begin{figure}
\includegraphics[scale=0.3]{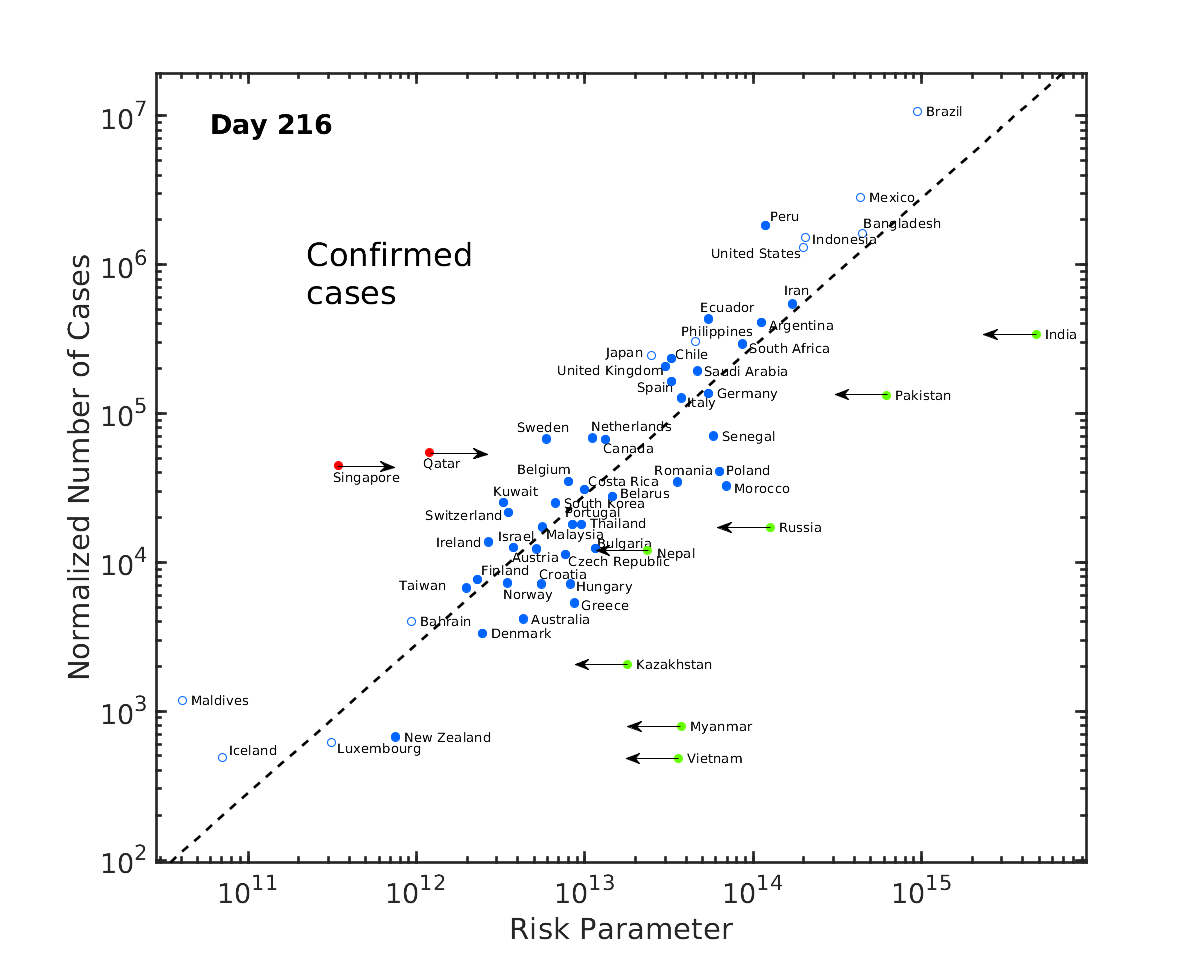}\caption{\label{fig:Outliers}The testing-normalized number of cases as a function
of the risk-parameter, including the expected outliers. China neighboring
countries (shown in green) could potentially receive considerable
ground traffic from China besides the flight-traffic, while only the
latter is included in the exposure parameter. The risk-parameters
of these countries are likely to be underestimated. Flight hubs (relative
to population size; shown in red) are likely to have considerable
incoming traffic from China of people who only transit through these
countries. The risk-parameter of these countries is likely to be overestimated.
The arrows emphasize that that the risk-parameters for hub-countries
only serve as lower-limits and that the risk-parameters of neighboring-countries
only serve as upper limits. }
\end{figure}

Although not included in the main analysis, given the uncertainties,
these countries do provide some indirect information. In Fig. \ref{fig:Outliers}
we show the same data as in Fig. \ref{fig:Correlation}, but now highlighting
the expected ``outlier'' underestimated exposure China-neighbors
and overestimated exposure flight-hubs countries. As can be seen the
locations of these countries and the outlying position are consistent
with the appropriate expectations, giving further (though indirect)
support for the double-virus co-infection model. In particular, in
a single HVS model, the neighbouring countries should be more exposed
and show \emph{larger} infection-levels and the flight-hubs should
be less-exposed showing \emph{lower} infection level, i.e. the opposite
expectation in respect to the double-strain model, and inconsistent
with the COVID-19 data. 

\subsection{Preceding low virulence strain and the co-infection SIR model}

\label{subsec:double-strain-model}

In order to demonstrate the basic aspects of the model and its observable
expectations, we make use of the Susceptible-Infected-Recovered (SIR)
model\cite{Ker+27}, often used to study the spread of infectious
disease. We extend it to the case of two circulating cross-immunizing
viruses (i.e. a person infected by one of the viruses is immune to
infection from the other virus, both during and after the infection),
following past studies of co-infection and the spread of multiple
strains\cite{Die+79,Cas+89,Adl+91,Bal+11,Sus+15,Kuc+16,Nic+16,Chi+20}),
as described by Eqs. \ref{eq:2a}-\ref{eq:2e} in the main text. 

Let us now consider the dynamics of the epidemic in two possible scenarios
(1) the HVS spreads in a given country earlier than the LVS, or the
LVS spreads earlier, but the HVS spreads in a given country shortly
after; and (2) The LVS spreads much earlier in a given country well
before the spread of the HVS. In the first scenario the HVS spreads
into the country before or shortly after the first arrival of the
LVS. In this case (see \ref{fig:model}b), although some fraction
of the population becomes LVS-infected, the spread of the HVS outruns
the LVS spread, even if the LVS began before (but not too long before,
as that would lead to the second scenario) and HVS effectively spreads
freely with little effect of the early LVS-spread. In particular,
in such cases the HVS-infection level could infect the majority of
the population, if no measures are taken, and until then its spread
is independent of the country population size, and just grows exponentially
with time, as if only a single strain exists. Note that the same evolution
also describes the early stages in the second scenario, before the
LVS reaches herd-immunity and affects the spread of the HVS. 

In these cases, the infection-level at a given time should just be
approximately 
\begin{equation}
{\rm I_{H}(t)=2^{(t-t_{exp})/T_{H}}},\label{eq:exponential-growth}
\end{equation}
 where $t_{exp}$ is the time of the initial infection (exposure)
in the given country, which is a function of the exposure level to
China, as we discuss below. The number of infected people arriving
to another destination country from China is proportional to the number
of infected people in China, and the fraction of those people who
travel from China to that destination. The number of infected people
in China at a given time (assuming no containment measures are taken)
is $2^{(t-t_{0})/T}$, where $t_{0}$ is the time of the initial infection
in China, and $\text{{\rm T}}$ is the doubling time of a given virus.
The fraction of travelers from China to a given country is proportional
to the mobility exposure level, $c_{exp}=\alpha\mu$ to that country,
where $\mu$ is taken from the GLEAM project (see section \ref{subsec:dat}),
and $\alpha$ is some constant unit calibration to account for the
GLEAM data units. The time for the first infection (exposure) in a
given country ($t_{exp})$ therefore follows the following relation 

\[
1=2^{(t_{exp}-t_{0})/T}c_{exp},
\]
which can be written as 
\begin{equation}
t_{exp}-t_{0}=-{\rm T}{\rm \ln_{2}}c_{exp}.\label{eq:t_exp}
\end{equation}

Plugging this result in Eq. \ref{eq:exponential-growth} we get
\begin{equation}
{\rm I_{H}(t)=2^{(t-t_{exp})/T_{H}}}=2^{(t-t_{0})/{\rm T}}c_{exp},\label{eq:free-exponential-growth}
\end{equation}

In other words, at a given time we expect to see a direct linear correlation
between the exposure level, and the number of confirmed cases at given
country, at least until the majority of the population became infected
and a herd immunity was achieved (by either virus). This is consistent
with the results in Fig. \ref{fig:Correlation}. At early time the
evolution follows the regular dynamics of a pandemic by a single virus,
and allows us to find the calibration parameter $\alpha$ by simple
linear regression on the data. At later times we see a different behavior,
which can be understood by considering the second scenario. 

In the second scenario, the LVS begins spreading in the community
and the number of LVS-infected people grows exponentially following
the same evolution as described for the HVS in Eq. \ref{eq:exponential-growth}.
If the LVS spreads far earlier than the HVS it could have infected
the majority of the population as to induce herd-immunity, and by
the time of first HVS infection, the HVS can not spread. If the HVS
begins its spread long after the LVS, but before the LVS achieved
herd-immunity, both viruses initially grow almost unimpeded, until
the LVS achieves herd-immunity, at which point both viruses spreads
stop the exponential growth, slow down and achieve their maximum (See
Fig. \ref{fig:model}). 

Longer delay between the initial spread of the LVS and the initial
spread of the HVS therefore translates into lower number of HVS infected
people and further constrains its spread up to a lower fraction of
the overall population (compare Figs. \ref{fig:model}b and d to Figs.
\ref{fig:model}a and c, respectively). For the same delay, countries
with smaller population sizes will achieve herd-immunity sooner, allowing
for a lesser number of HVS infections (compare Figs. \ref{fig:model}1
and b to Figs. \ref{fig:model}c and d, respectively).

The introduction of a precursor LVS can therefore explain the otherwise
puzzling findings of the highest exposed countries showing low infection
levels at late times (Fig. \ref{fig:Correlation}a), and the transition
from initial fast exponential growth to a slow non-expoential level,
and even an effective complete stop of the growth in all countries,
although \emph{none} of them have infection levels even close to those
required for herd-immunity. In fact, all countries show far lower
infection levels than would be expected from a simple exponential
growth expected from Eq. \ref{eq:exponential-growth} (Fig. \ref{fig:cases-vs-time}).

Given some initial number of LVS-infected people (${\rm I_{L}^{0}}$)
in the population at the point when the HVS begins to spread (i.e.
the time of first HVS-infection), in a given country, one can find
specific relations between the maximal number of infected people,
the doubling times $T_{L}$ and $T_{H}$ of the two viruses, and the
initial fraction of LVS and HVS infected people in the population.
During the initial free spread of the LVS, it can spread until it
achieves herd-immunity once the majority of the population is infected
(we will take the actual population size for simplicity). The time
it takes to the maximal infection point is therefore the number of
doublings of the infected population from its initial value up to
approximately the size of the population $\text{\ensuremath{\sim}}\text{N}$,
times the doubling time
\begin{equation}
t_{max}=T_{L}\cdot{\rm log_{2}}\left(\frac{N}{I_{L}^{0}}\right)\label{eq:t_max}
\end{equation}

Until the LVS achieves herd-immunity, the HVS spreads freely. Its
spread should follow an exponential growth until slowed/stopped once
herd immunity is achieved. As discussed above, if the LVS infects
the majority of the population and achieves herd-immunity before being
outrun by the spread of the HVS, the mutual immunity insures that
the further spread of the HVS will also be quenched. At this8point
the HVS would also stop and achieve its maximum infection level. Hence,
the number of HVS-infected people is just

\begin{equation}
{\rm I}_{{\rm H(t_{max})}}={\rm {\rm I}_{H}^{max}}{\rm =}2^{t_{max}/T_{{\rm H}}}=2^{\left({\rm \frac{T_{L}}{T_{H}}}\right){\rm log_{2}}\left({\rm \frac{N}{I_{L}^{0}}}\right)}=\left({\rm \frac{N}{I_{L}^{0}}}\right)^{{\rm T_{L}/T_{H}}},\label{eq:IH_max}
\end{equation}
or writing it in logarithmic terms

\begin{equation}
{\rm log_{2}\left(I_{{\rm H}}^{max}\right)}=-\left(\frac{T_{L}}{T_{H}}\right){\rm log_{2}}\left(\frac{I_{L}^{0}}{N}\right).\label{eq:Finding Td_L}
\end{equation}

The exposure time to a virus for a given country was derived in Eq.
\ref{eq:t_exp}. It is valid both for the LVS and the HVS, where the
parameters correspond to the specific strain (i.e. $t_{exp}^{L},\,t_{0}^{L}$
and $T_{L}$ for the LVS, and $t_{exp}^{H},\,t_{0}^{H}$ and $T_{H}$
for the HVS, and ${\rm c}_{exp}$ is the same for both (neglecting
seasonal variations in mobility). Using this relation, we can find
the infection level of the LVS at a given country at the point when
the HVS first arrives, ${\rm I_{L}^{0}}$. We get 

\[
t_{exp}^{L}-t_{0}^{L}=-T_{L}{\rm \ln_{2}}c,({\rm country)}
\]

\[
t_{exp}^{H}-t_{0}^{H}=-T_{H}{\rm \ln_{2}}c,({\rm country)}
\]

and therefore
\begin{equation}
t_{exp}^{H}-t_{exp}^{L}=(t_{0}^{H}-t_{0}^{L})-(T_{H}-T_{L})\ln_{2}(c_{exp}).\label{eq:time-difference}
\end{equation}

Plugging this into Eq. \ref{eq:IH_max} we get

\[
I_{L}^{0}=2^{(t_{exp}^{H}-t_{exp}^{L})/T_{L}}=2^{\left[t_{0}^{H}-t_{0}^{L}-(T_{H}^{d}-T_{L}^{d})\ln_{2}(c_{exp})\right]/T_{L}}=2^{(t_{0}^{H}-t_{0}^{L})/T_{L}}c_{exp}^{1-T_{H}/T_{L}},
\]
to finally get from Eq. \ref{eq:Finding Td_L} 

\[
{\rm I}_{{\rm H}}^{max}=\left({\rm \frac{N}{I_{L}^{0}}}\right)^{{\rm T_{L}/T_{H}}}=\left(\frac{{\rm N}}{{\rm 2^{(t_{0}^{H}-t_{0}^{L})/T_{L}}c_{exp}{}^{1-T_{H}/T_{L}}}}\right)^{{\rm T_{L}/T_{H}}},
\]
or in logarithmic terms

\begin{equation}
\log_{2}({\rm I}_{{\rm H}}^{{\rm max}})=[(t_{0}^{L}-t_{0}^{H})/{\rm T}_{{\rm H}}]+\left(\frac{{\rm T_{L}}}{{\rm T_{H}}}\right)\log_{2}{\rm N}-\left(\frac{{\rm T_{L}}}{{\rm T_{H}}}-1\right)\log_{2}{\rm c}_{exp},\label{eq:IHmax-log}
\end{equation}
i.e. the \emph{maximal} number of infected people in a given country
in which the LVS reached herd-immunity before the HVS, should be directly
correlated with the total population in the country, and inversely
correlated with the exposure level to China. The exact powers are
determined by the doubling times (spread rates) of the LVS and the
HVS. The larger the exposure to China the longer the delay between
the initial LVS infection and the initial HVS infection (see Eq. \ref{eq:time-difference}),
allowing for the LVS more time to spread and achieve herd-immunity
while the HVS is still at earlier stages. The smaller the population
size in a given country, the faster the LVS can achieve herd-immunity,
and quench the spread of the LVS at earlier times, leading to lower
infection levels, consistent with the models in Fig. \ref{fig:model}.
A critical transition can occur for less exposed and/or more populated
countries, when ${\rm I}_{{\rm H}}^{{\rm max}}$ is comparable with
the overall population, i.e. providing no constraint on the HVS spread
(needless to say ${\rm I}_{{\rm H}}^{{\rm max}}$ needs always to
be smaller or equal to the total population ${\rm N})$. This happens
when the HVS had sufficient time to spread and eventually outrun the
spread of the LVS, before the latter achieved herd-immunity level.
As discussed in the main text, we can therefore fit our model using
the data on the infection-level (testing-normalized cases or age-normalized
deaths) and the mobility-exposure data, to find the model parameters.
In particular, we can infer the doubling-time of the LVS using a linear-regression
analysis, and then also infer the delay between the onset of the LVS
and the HVS (see main text), which then also provides the mobility-exposure
calibration unit $\alpha$ mentioned above.

For any given country we can now predict the expected infection levels.
Our models in Figs. \ref{fig:model} suggest that the HVS spread should
already slow down when the infection level reaches $\sim10\%$ of
the maximal level, and that the rise and decay of the HVS infection
level should be asymmetric with a fast rise and a slow decay, in particular
it is more asymmetric than would be expected for the case of a single
virus evolution (e.g. compare the more symmetric structure of the
LVS evolution in comparison with the HVS), potentially consistent
with observed asymmetries in the epidemic dynamics observed in many
countries. 

\subsection{Dynamics of early spread}

\begin{figure}
\includegraphics[scale=0.3]{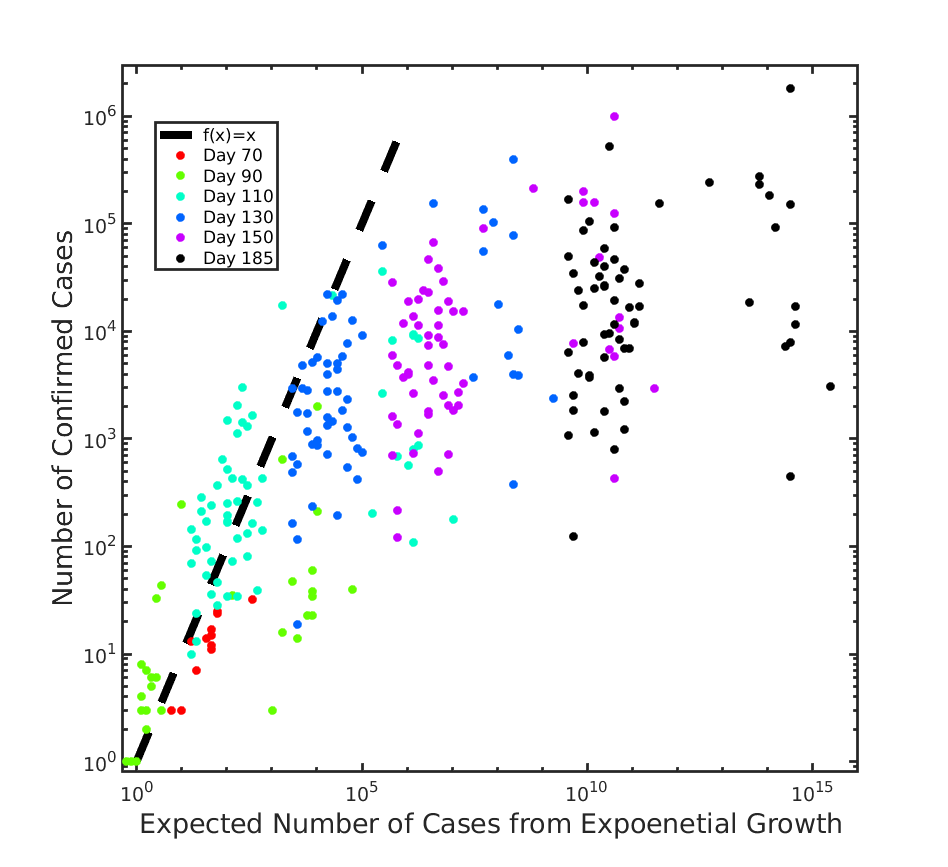}\caption{\label{fig:cases-vs-time}The number of confirmed cases vs. the expected
number of cases for an exponential growth, at epochs. Early on (day
70) all infected countries followed a simple exponential growth (compare
with the dashed line). At later times newly infected countries follow
an exponential growth, while countries infected earlier stop following
an exponential growth and show a sub-exponential growth. This patterns
repeats, with more countries stopping their exponential growth after
their initial exponential spread. By day 185 all countries in our
sample already stopped spreading exponentially. This dynamics follows
the expectation from the double-strain co-infection model (see text).
The dashed line corresponds to a simple exponential growth with a
doubling time of 2.7 days.}
 
\end{figure}

If allowed to spread with no constraint the number of infected people
should rise exponentially to be

\begin{equation}
{\rm log}_{2}\left[{\rm I_{exp}}(t)\right]=(t-t_{exp})/{\rm T},\label{eq:exponential-growth-1}
\end{equation}
at time $t$, where ${\rm I}$ is the number of infected people, $t_{0}$
is the time of the initial exposure (first infection), and ${\rm T}$
is the doubling time for the the given virus. As discussed above (section
\ref{subsec:double-strain-model}), in the early stages of the HVS
infection the evolution should effectively be indistinguishable from
that expected for a the unconstrained spread of a single virus. In
Fig. \ref{fig:cases-vs-time} we show the number of age-corrected
deaths in different countries at several points in time, as a function
of the initial infection time in each country. As expected, in countries
infected earlier (larger number of days since first infection), the
number of age-normalized deaths is larger, and consistent with an
exponential growth as featured in Eq. \ref{eq:exponential-growth-1}.
However, at later times one can observe that the number of deaths
stops growing exponentially, and the overall structure of the number
of deaths in each country hardly evolves (on logarithmic scales; one
can still see a very slow, non-exponential growth). This behavior
is explained by the double-strain co-infection model, which suggests
(see Fig. \ref{fig:model}) that the exponential growth should begin
to slow down once the LVS-infection levels reach a large fraction
of the population, and later become sub-exponential until it eventually
stops growing. 

\bibliographystyle{naturemag}

\begin{thebibliography}{10}
\expandafter\ifx\csname url\endcsname\relax
  \def\url#1{\texttt{#1}}\fi
\expandafter\ifx\csname urlprefix\endcsname\relax\def\urlprefix{URL }\fi
\providecommand{\bibinfo}[2]{#2}
\providecommand{\eprint}[2][]{\url{#2}}

\bibitem{Lu+20}
\bibinfo{author}{Lu, R.} \emph{et~al.}
\newblock \bibinfo{title}{{{G}enomic characterisation and epidemiology of 2019
  novel coronavirus: implications for virus origins and receptor binding}}.
\newblock \emph{\bibinfo{journal}{Lancet}} \textbf{\bibinfo{volume}{395}},
  \bibinfo{pages}{565--574} (\bibinfo{year}{2020}).

\bibitem{Wu+20}
\bibinfo{author}{Wu, A.} \emph{et~al.}
\newblock \bibinfo{title}{{{G}enome {C}omposition and {D}ivergence of the
  {N}ovel {C}oronavirus (2019-n{C}o{V}) {O}riginating in {C}hina}}.
\newblock \emph{\bibinfo{journal}{Cell Host Microbe}}
  \textbf{\bibinfo{volume}{27}}, \bibinfo{pages}{325--328}
  (\bibinfo{year}{2020}).

\bibitem{Hai+20}
\bibinfo{author}{Haider, N.} \emph{et~al.}
\newblock \bibinfo{title}{{{P}assengers' destinations from {C}hina: low risk of
  {N}ovel {C}oronavirus (2019-n{C}o{V}) transmission into {A}frica and {S}outh
  {A}merica}}.
\newblock \emph{\bibinfo{journal}{Epidemiol. Infect.}}
  \textbf{\bibinfo{volume}{148}}, \bibinfo{pages}{e41} (\bibinfo{year}{2020}).

\bibitem{Van+11}
\bibinfo{author}{Van~den Broeck, W.} \emph{et~al.}
\newblock \bibinfo{title}{{{T}he {G}{L}{E}a{M}viz computational tool, a
  publicly available software to explore realistic epidemic spreading scenarios
  at the global scale}}.
\newblock \emph{\bibinfo{journal}{BMC Infect. Dis.}}
  \textbf{\bibinfo{volume}{11}}, \bibinfo{pages}{37} (\bibinfo{year}{2011}).

\bibitem{Pas+19}
\bibinfo{author}{Pastore~y Piontti, A.}, \bibinfo{author}{Perra, N.},
  \bibinfo{author}{Rossi, L.}, \bibinfo{author}{Samay, N.} \&
  \bibinfo{author}{Vespignani, A.}
\newblock \emph{\bibinfo{title}{Charting the Next Pandemic: Modeling Infectious
  Disease Spreading in the Data Science Age}} (\bibinfo{year}{2019}).

\bibitem{Chi+20}
\bibinfo{author}{Chinazzi, M.} \emph{et~al.}
\newblock \bibinfo{title}{The effect of travel restrictions on the spread of
  the 2019 novel coronavirus (covid-19) outbreak}.
\newblock \emph{\bibinfo{journal}{Science}} \textbf{\bibinfo{volume}{368}},
  \bibinfo{pages}{395--400} (\bibinfo{year}{2020}).
\newblock \urlprefix\url{https://science.sciencemag.org/content/368/6489/395}.
\newblock
  \eprint{https://science.sciencemag.org/content/368/6489/395.full.pdf}.

\bibitem{Die+79}
\bibinfo{author}{Dietz, K.}
\newblock \bibinfo{title}{{{E}pidemiologic interference of virus populations}}.
\newblock \emph{\bibinfo{journal}{J Math Biol}} \textbf{\bibinfo{volume}{8}},
  \bibinfo{pages}{291--300} (\bibinfo{year}{1979}).

\bibitem{Cas+89}
\bibinfo{author}{Castillo-Chavez, C.}, \bibinfo{author}{Hethcote, H.~W.},
  \bibinfo{author}{Andreasen, V.}, \bibinfo{author}{Levin, S.~A.} \&
  \bibinfo{author}{Liu, W.~M.}
\newblock \bibinfo{title}{{{E}pidemiological models with age structure,
  proportionate mixing, and cross-immunity}}.
\newblock \emph{\bibinfo{journal}{J Math Biol}} \textbf{\bibinfo{volume}{27}},
  \bibinfo{pages}{233--258} (\bibinfo{year}{1989}).

\bibitem{Adl+91}
\bibinfo{author}{Adler, F.~R.} \& \bibinfo{author}{Brunet, R.~C.}
\newblock \bibinfo{title}{{{T}he dynamics of simultaneous infections with
  altered susceptibilities}}.
\newblock \emph{\bibinfo{journal}{Theor Popul Biol}}
  \textbf{\bibinfo{volume}{40}}, \bibinfo{pages}{369--410}
  (\bibinfo{year}{1991}).

\bibitem{Bal+11}
\bibinfo{author}{Balmer, O.} \& \bibinfo{author}{Tanner, M.}
\newblock \bibinfo{title}{{{P}revalence and implications of multiple-strain
  infections}}.
\newblock \emph{\bibinfo{journal}{Lancet Infect Dis}}
  \textbf{\bibinfo{volume}{11}}, \bibinfo{pages}{868--878}
  (\bibinfo{year}{2011}).

\bibitem{Sus+15}
\bibinfo{author}{Susi, H.}, \bibinfo{author}{Barr?s, B.},
  \bibinfo{author}{Vale, P.~F.} \& \bibinfo{author}{Laine, A.~L.}
\newblock \bibinfo{title}{{{C}o-infection alters population dynamics of
  infectious disease}}.
\newblock \emph{\bibinfo{journal}{Nat Commun}} \textbf{\bibinfo{volume}{6}},
  \bibinfo{pages}{5975} (\bibinfo{year}{2015}).

\bibitem{Kuc+16}
\bibinfo{author}{Kucharski, A.~J.}, \bibinfo{author}{Andreasen, V.} \&
  \bibinfo{author}{Gog, J.~R.}
\newblock \bibinfo{title}{{{C}apturing the dynamics of pathogens with many
  strains}}.
\newblock \emph{\bibinfo{journal}{J Math Biol}} \textbf{\bibinfo{volume}{72}},
  \bibinfo{pages}{1--24} (\bibinfo{year}{2016}).

\bibitem{Nic+16}
\bibinfo{author}{Nickbakhsh, S.} \emph{et~al.}
\newblock \bibinfo{title}{{{M}odelling the impact of co-circulating low
  pathogenic avian influenza viruses on epidemics of highly pathogenic avian
  influenza in poultry}}.
\newblock \emph{\bibinfo{journal}{Epidemics}} \textbf{\bibinfo{volume}{17}},
  \bibinfo{pages}{27--34} (\bibinfo{year}{2016}).

\bibitem{Tho+19}
\bibinfo{author}{Thompson, R.}, \bibinfo{author}{Thompson, C.},
  \bibinfo{author}{Pelerman, O.}, \bibinfo{author}{Gupta, S.} \&
  \bibinfo{author}{Obolski, U.}
\newblock \bibinfo{title}{Increased frequency of travel in the presence of
  cross-immunity may act to decrease the chance of a global pandemic}.
\newblock \emph{\bibinfo{journal}{Phil. Trans. R. Soc. B}}
  \textbf{\bibinfo{volume}{374}}, \bibinfo{pages}{20180274}
  (\bibinfo{year}{2019}).
\newblock
  \urlprefix\url{https://www.biorxiv.org/content/early/2019/01/18/404871}.
\newblock
  \eprint{https://www.biorxiv.org/content/early/2019/01/18/404871.full.pdf}.

\bibitem{Kam+20}
\bibinfo{author}{Kamikubo, Y.} \& \bibinfo{author}{Takahashi, A.}
\newblock \bibinfo{title}{Epidemiological tools that predict partial herd
  immunity to sars coronavirus 2}.
\newblock \emph{\bibinfo{journal}{medRxiv}}  (\bibinfo{year}{2020}).
\newblock
  \urlprefix\url{https://www.medrxiv.org/content/early/2020/03/27/2020.03.25.20043679}.
\newblock
  \eprint{https://www.medrxiv.org/content/early/2020/03/27/2020.03.25.20043679.full.pdf}.

\bibitem{Qi+2018}
\bibinfo{author}{Qi, W.} \emph{et~al.}
\newblock \bibinfo{title}{{{E}mergence and {A}daptation of a {N}ovel {H}ighly
  {P}athogenic {H}7{N}9 {I}nfluenza {V}irus in {B}irds and {H}umans from a 2013
  {H}uman-{I}nfecting {L}ow-{P}athogenic {A}ncestor}}.
\newblock \emph{\bibinfo{journal}{J. Virol.}} \textbf{\bibinfo{volume}{92}}
  (\bibinfo{year}{2018}).

\bibitem{Seo+01}
\bibinfo{author}{Seo, S.~H.} \& \bibinfo{author}{Webster, R.~G.}
\newblock \bibinfo{title}{{{C}ross-reactive, cell-mediated immunity and
  protection of chickens from lethal {H}5{N}1 influenza virus infection in
  {H}ong {K}ong poultry markets}}.
\newblock \emph{\bibinfo{journal}{J. Virol.}} \textbf{\bibinfo{volume}{75}},
  \bibinfo{pages}{2516--2525} (\bibinfo{year}{2001}).

\bibitem{Kha+09}
\bibinfo{author}{Khalenkov, A.}, \bibinfo{author}{Perk, S.},
  \bibinfo{author}{Panshin, A.}, \bibinfo{author}{Golender, N.} \&
  \bibinfo{author}{Webster, R.~G.}
\newblock \bibinfo{title}{{{M}odulation of the severity of highly pathogenic
  {H}5{N}1 influenza in chickens previously inoculated with {I}sraeli {H}9{N}2
  influenza viruses}}.
\newblock \emph{\bibinfo{journal}{Virology}} \textbf{\bibinfo{volume}{383}},
  \bibinfo{pages}{32--38} (\bibinfo{year}{2009}).

\bibitem{Nfo+12}
\bibinfo{author}{Nfon, C.} \emph{et~al.}
\newblock \bibinfo{title}{{{P}rior infection of chickens with {H}1{N}1 or
  {H}1{N}2 avian influenza elicits partial heterologous protection against
  highly pathogenic {H}5{N}1}}.
\newblock \emph{\bibinfo{journal}{PLoS ONE}} \textbf{\bibinfo{volume}{7}},
  \bibinfo{pages}{e51933} (\bibinfo{year}{2012}).

\bibitem{Li+20b}
\bibinfo{author}{Li, X.} \emph{et~al.}
\newblock \bibinfo{title}{{{T}ransmission dynamics and evolutionary history of
  2019-n{C}o{V}}}.
\newblock \emph{\bibinfo{journal}{J. Med. Virol.}}
  \textbf{\bibinfo{volume}{92}}, \bibinfo{pages}{501--511}
  (\bibinfo{year}{2020}).

\bibitem{Su+15}
\bibinfo{author}{Su, Y. C.~F.} \emph{et~al.}
\newblock \bibinfo{title}{{{P}hylodynamics of {H}1{N}1/2009 influenza reveals
  the transition from host adaptation to immune-driven selection}}.
\newblock \emph{\bibinfo{journal}{Nat Commun}} \textbf{\bibinfo{volume}{6}},
  \bibinfo{pages}{7952} (\bibinfo{year}{2015}).

\bibitem{Fer+03}
\bibinfo{author}{Ferguson, N.~M.}, \bibinfo{author}{Galvani, A.~P.} \&
  \bibinfo{author}{Bush, R.~M.}
\newblock \bibinfo{title}{{{E}cological and immunological determinants of
  influenza evolution}}.
\newblock \emph{\bibinfo{journal}{Nature}} \textbf{\bibinfo{volume}{422}},
  \bibinfo{pages}{428--433} (\bibinfo{year}{2003}).

\bibitem{Nic+19}
\bibinfo{author}{Nickbakhsh, S.} \emph{et~al.}
\newblock \bibinfo{title}{{{V}irus-virus interactions impact the population
  dynamics of influenza and the common cold}}.
\newblock \emph{\bibinfo{journal}{Proc. Natl. Acad. Sci. U.S.A.}}
  (\bibinfo{year}{2019}).

\bibitem{Che+20}
\bibinfo{author}{Chen, Z.-w.}, \bibinfo{author}{Li, Z.}, \bibinfo{author}{Li,
  H.}, \bibinfo{author}{Ren, H.} \& \bibinfo{author}{Hu, P.}
\newblock \bibinfo{title}{Global genetic diversity patterns and transmissions
  of sars-cov-2}.
\newblock \emph{\bibinfo{journal}{medRxiv}}  (\bibinfo{year}{2020}).
\newblock
  \urlprefix\url{https://www.medrxiv.org/content/early/2020/05/08/2020.05.05.20091413}.
\newblock
  \eprint{https://www.medrxiv.org/content/early/2020/05/08/2020.05.05.20091413.full.pdf}.

\bibitem{Yao+20}
\bibinfo{author}{Yao, H.} \emph{et~al.}
\newblock \bibinfo{title}{Patient-derived mutations impact pathogenicity of
  sars-cov-2}.
\newblock \emph{\bibinfo{journal}{medRxiv}}  (\bibinfo{year}{2020}).
\newblock
  \urlprefix\url{https://www.medrxiv.org/content/early/2020/04/22/2020.04.14.20060160}.
\newblock
  \eprint{https://www.medrxiv.org/content/early/2020/04/22/2020.04.14.20060160.full.pdf}.

\bibitem{Urb+16}
\bibinfo{author}{Urbanowicz, R.~A.} \emph{et~al.}
\newblock \bibinfo{title}{{{H}uman {A}daptation of {E}bola {V}irus during the
  {W}est {A}frican {O}utbreak}}.
\newblock \emph{\bibinfo{journal}{Cell}} \textbf{\bibinfo{volume}{167}},
  \bibinfo{pages}{1079--1087} (\bibinfo{year}{2016}).

\bibitem{Die+16}
\bibinfo{author}{Diehl, W.~E.} \emph{et~al.}
\newblock \bibinfo{title}{{{E}bola {V}irus {G}lycoprotein with {I}ncreased
  {I}nfectivity {D}ominated the 2013-2016 {E}pidemic}}.
\newblock \emph{\bibinfo{journal}{Cell}} \textbf{\bibinfo{volume}{167}},
  \bibinfo{pages}{1088--1098} (\bibinfo{year}{2016}).

\bibitem{Het00}
\bibinfo{author}{Hethcote, H.~W.}
\newblock \bibinfo{title}{The mathematics of infectious diseases}.
\newblock \emph{\bibinfo{journal}{SIAM Review}} \textbf{\bibinfo{volume}{42}},
  \bibinfo{pages}{599--653} (\bibinfo{year}{2000}).

\bibitem{MATLAB}
\bibinfo{author}{MATLAB}.
\newblock \emph{\bibinfo{title}{version 7.10.0 (R2010a)}}
  (\bibinfo{publisher}{The MathWorks Inc.}, \bibinfo{address}{Natick,
  Massachusetts}, \bibinfo{year}{2010}).

\bibitem{Ker+27}
\bibinfo{author}{Kermack, W.~O.}, \bibinfo{author}{McKendrick, A.~G.},
  \bibinfo{author}{Kermack, W.~O.} \& \bibinfo{author}{McKendrick, A.~G.}
\newblock \bibinfo{title}{{{C}ontributions to the mathematical theory of
  epidemics--{I}. 1927}}.
\newblock \emph{\bibinfo{journal}{Bull. Math. Biol.}}
  \textbf{\bibinfo{volume}{53}}, \bibinfo{pages}{33--55}
  (\bibinfo{year}{1991}).

\bibitem{Bri+20}
\bibinfo{author}{Britton, T.}, \bibinfo{author}{Trapman, P.} \&
  \bibinfo{author}{Ball, F.~G.}
\newblock \bibinfo{title}{The disease-induced herd immunity level for covid-19
  is substantially lower than the classical herd immunity level}.
\newblock \emph{\bibinfo{journal}{medRxiv}}  (\bibinfo{year}{2020}).
\newblock
  \urlprefix\url{https://www.medrxiv.org/content/early/2020/05/14/2020.05.06.20093336}.
\newblock
  \eprint{https://www.medrxiv.org/content/early/2020/05/14/2020.05.06.20093336.full.pdf}.

\bibitem{Ioa+20}
\bibinfo{author}{Ioannidis, J.}
\newblock \bibinfo{title}{The infection fatality rate of covid-19 inferred from
  seroprevalence data}.
\newblock \emph{\bibinfo{journal}{medRxiv}}  (\bibinfo{year}{2020}).
\newblock
  \urlprefix\url{https://www.medrxiv.org/content/early/2020/05/19/2020.05.13.20101253}.
\newblock
  \eprint{https://www.medrxiv.org/content/early/2020/05/19/2020.05.13.20101253.full.pdf}.

\bibitem{Hav+20}
\bibinfo{author}{Havers, F.~P.} \emph{et~al.}
\newblock \bibinfo{title}{Seroprevalence of antibodies to sars-cov-2 in six
  sites in the united states, march 23-may 3, 2020}.
\newblock \emph{\bibinfo{journal}{medRxiv}}  (\bibinfo{year}{2020}).
\newblock
  \urlprefix\url{https://www.medrxiv.org/content/early/2020/06/26/2020.06.25.20140384}.
\newblock
  \eprint{https://www.medrxiv.org/content/early/2020/06/26/2020.06.25.20140384.full.pdf}.

\bibitem{Ste+05}
\bibinfo{author}{Stephenson, I.} \emph{et~al.}
\newblock \bibinfo{title}{{{C}ross-reactivity to highly pathogenic avian
  influenza {H}5{N}1 viruses after vaccination with nonadjuvanted and
  {M}{F}59-adjuvanted influenza {A}/{D}uck/{S}ingapore/97 ({H}5{N}3) vaccine: a
  potential priming strategy}}.
\newblock \emph{\bibinfo{journal}{J. Infect. Dis.}}
  \textbf{\bibinfo{volume}{191}}, \bibinfo{pages}{1210--1215}
  (\bibinfo{year}{2005}).

\bibitem{Man+20}
\bibinfo{author}{Manenti, A.} \emph{et~al.}
\newblock \bibinfo{title}{{{E}valuation of {S}{A}{R}{S}-{C}o{V}-2 neutralizing
  antibodies using a {C}{P}{E}-based colorimetric live virus
  micro-neutralization assay in human serum samples}}.
\newblock \emph{\bibinfo{journal}{J. Med. Virol.}}  (\bibinfo{year}{2020}).

\bibitem{Sek+20}
\bibinfo{author}{Sekine, T.} \emph{et~al.}
\newblock \bibinfo{title}{Robust t cell immunity in convalescent individuals
  with asymptomatic or mild covid-19}.
\newblock \emph{\bibinfo{journal}{bioRxiv}}  (\bibinfo{year}{2020}).
\newblock
  \urlprefix\url{https://www.biorxiv.org/content/early/2020/06/29/2020.06.29.174888}.
\newblock
  \eprint{https://www.biorxiv.org/content/early/2020/06/29/2020.06.29.174888.full.pdf}.

\bibitem{Fon+20}
\bibinfo{author}{Fongaro, G.} \emph{et~al.}
\newblock \bibinfo{title}{Sars-cov-2 in human sewage in santa catalina, brazil,
  november 2019}.
\newblock \emph{\bibinfo{journal}{medRxiv}}  (\bibinfo{year}{2020}).
\newblock
  \urlprefix\url{https://www.medrxiv.org/content/early/2020/06/29/2020.06.26.20140731}.
\newblock
  \eprint{https://www.medrxiv.org/content/early/2020/06/29/2020.06.26.20140731.full.pdf}.

\bibitem{Hua+20}
\bibinfo{author}{Huang, A.~T.} \emph{et~al.}
\newblock \bibinfo{title}{{{A} systematic review of antibody mediated immunity
  to coronaviruses: antibody kinetics, correlates of protection, and
  association of antibody responses with severity of disease}}.
\newblock \emph{\bibinfo{journal}{medRxiv}}  (\bibinfo{year}{2020}).

\bibitem{Hal+20}
\bibinfo{author}{Hale, N., T.}, \bibinfo{author}{Angrist, B.~K.},
  \bibinfo{author}{Petherick, A.}, \bibinfo{author}{Phillips, T.} \&
  \bibinfo{author}{Webster, S.}
\newblock \bibinfo{title}{Variations in government responses to covid-19}.
\newblock \emph{\bibinfo{journal}{Blavatnik School of Government, University of
  Oxford}} \textbf{\bibinfo{volume}{BSG-WP-2020/032}} (\bibinfo{year}{2020}).

\end{thebibliography}

\end{document}